\documentclass[12pt,preprintnumbers, nofootinbib]{revtex4-1}
\usepackage{subfigure,graphicx,amsmath}
\usepackage{multirow}

\begin{document}

\title{Diagnosing the top-quark angular asymmetry\\ using LHC intrinsic charge asymmetries}

\author{Simon Knapen}
\email{knapen@physics.rutgers.edu}
\affiliation{Department of Physics and Astronomy,
Rutgers University, Piscataway, NJ 08854}

\author{Yue Zhao}
\email{zhaoyue@physics.rutgers.edu}
\affiliation{Department of Physics and Astronomy,
Rutgers University, Piscataway, NJ 08854}

\author{Matthew J. Strassler}
\email{strassler@physics.rutgers.edu}
\affiliation{Department of Physics and Astronomy,
Rutgers University, Piscataway, NJ 08854}

\preprint{RUNHETC-2011-23}

\begin{abstract}
Flavor-violating interactions involving new heavy particles are
among proposed explanations for the $t\bar t$ forward-backward
asymmetry observed at the Tevatron. Many of these models generate a
$t\bar t$-plus-jet signal at the LHC.  In this paper we identify
several new charge asymmetric variables in $t\bar tj$ events that
can contribute to the discovery of such models at the LHC. We
propose a data-driven method for the background, largely eliminating
the need for a Monte Carlo prediction of $t\bar t$-plus-jets, and
thus reducing systematic errors.  With a fast detector simulation,
we estimate the statistical sensitivity of our variables for one of
these models, finding that charge-asymmetric variables could
materially assist in the exclusion of  the Standard Model across much of the mass and coupling range, given 5 inverse fb of data.  Should any signal appear, our variables will be
useful in distinguishing classes of models from one another.
\end{abstract}

\maketitle
\section{Introduction}
The most peculiar among the Standard Model fermions, the top quark has
challenged the high energy physics community, both on the experimental
and theoretical level, since its discovery in 1995. From the
theoretical viewpoint, its exceptional mass suggests that it might
play a special role in the mechanism of electroweak symmetry
breaking. This occurs in a number of proposed theories, including
Little Higgs and Top-color Assisted Technicolor, and even within
many supersymmetric models. On the experimental side, the
predictions of the Standard Model (SM) for the top quark are still not
fully tested.  At the Tevatron, the high production threshold limited the number of
$t\bar t$ events, and only now at
the LHC will it be possible to perform precision measurements of the
top quark's properties.

While most aspects of the top quark agree so far with SM
predictions, both the CDF \cite{Aaltonen:2011kc,CDFLeptons} and D0
\cite{Abazov:2011rq, :2007qb} collaborations have reported an
anomalous forward-backward asymmetry for $t\bar t$ pairs at
intermediate to high invariant mass, much larger than expected
from SM calculations
\cite{Kuhn:1998jr,Kuhn:1998kw,Bowen:2005ap,Almeida:2008ug,Ahrens:2011uf,Kuhn:2011ri}.
This result, which relies upon ``forward'' being defined relative to
the Tevatron's proton beam, cannot be immediately checked at a
proton-proton collider such as the LHC. However, it is well-known
that forward-backward asymmetries at a proton-antiproton machine
lead to differential charge asymmetries at a proton-proton machine,
and indeed, a differential charge asymmetry in $t\bar t$ production,
as a function of the $t$ quark's rapidity, should be observable.
This quantity has been discussed by theorists, for instance in
\cite{Langacker:1986,Dittmar:1996my,Li:2009xh,Wang:2010tg,Jung:2011zv,Hewett:2011wz},
and has been measured at the LHC experiments
\cite{ATLASChargeAsymm, CMSChargeAsymmearly, CMSChargeAsymm}.  The statistical errors on
this measurement are still rather large, however, and meanwhile the
LHC's higher energy allows its experiments to probe for related
phenomena in other ways.

\begin{figure}[t]
\begin{minipage}{0.6\textwidth}
\subfigure[ s-channel]{
   \includegraphics[scale =0.3] {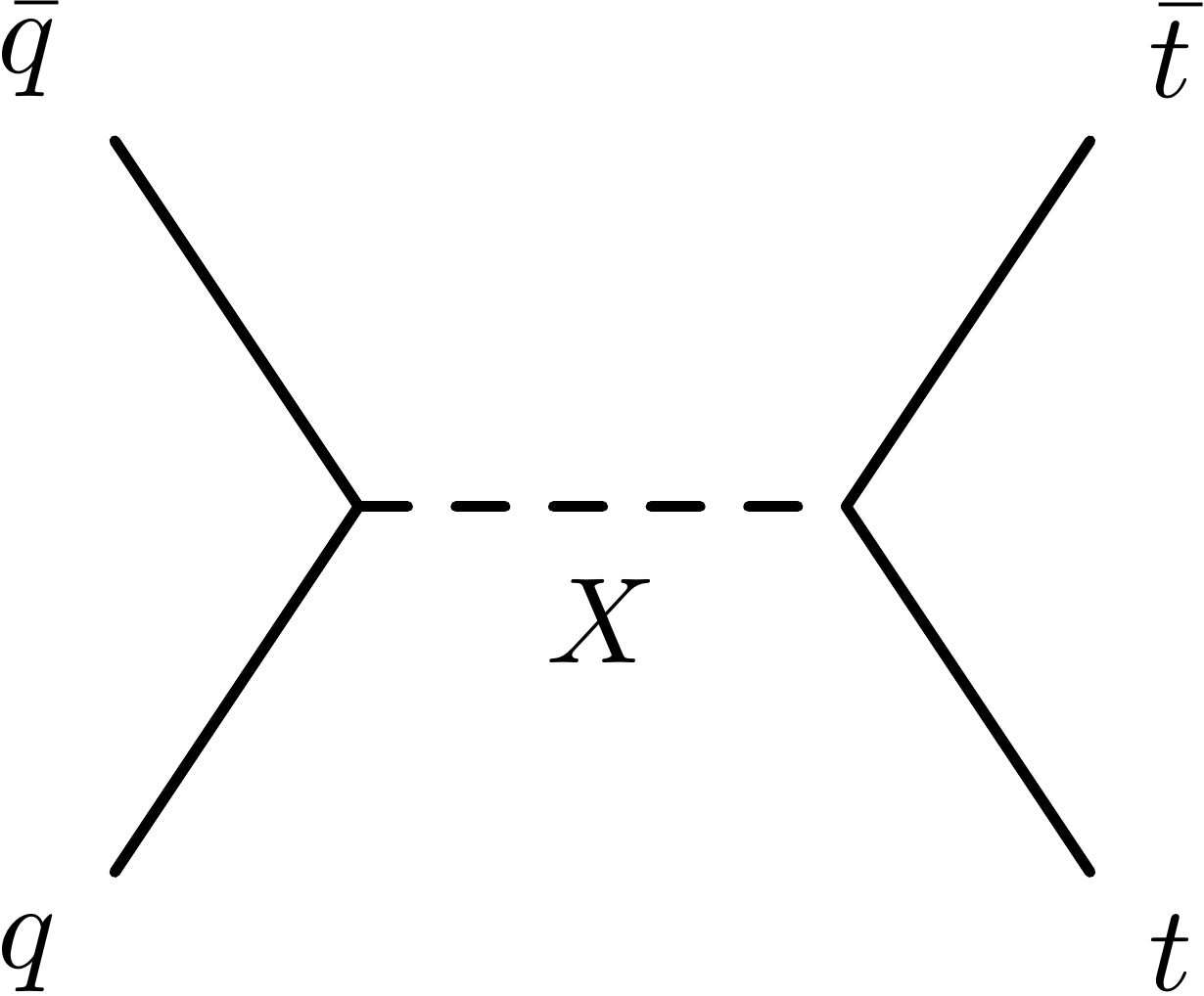}
   \label{Xschannel}
 }\hfill
 \subfigure[ t-channel]{
   \includegraphics[scale =0.3] {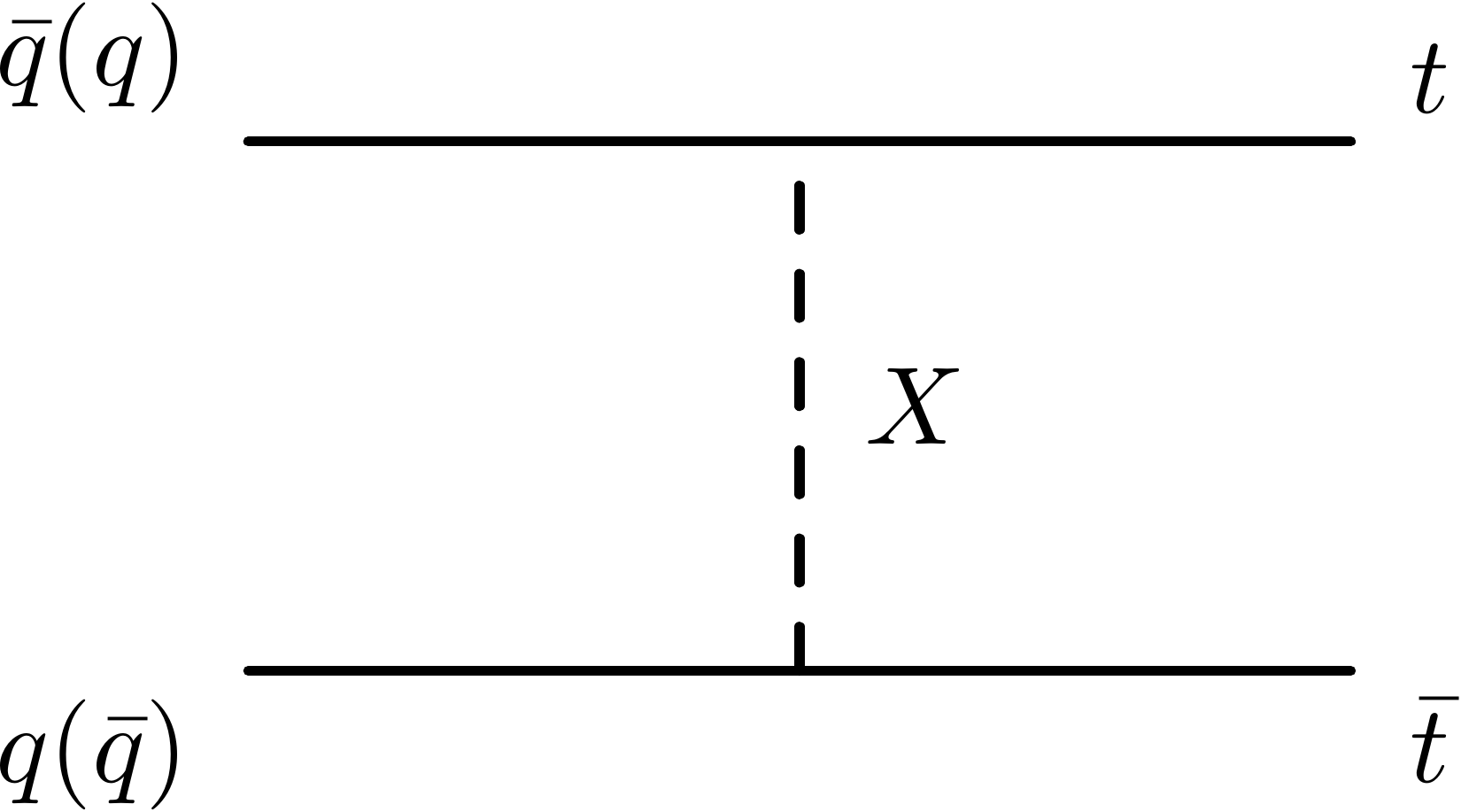}
   \label{Xtchannel}
 }
\caption{Diagrams that can lead to a forward-backward asymmetry at the Tevatron in $t\bar t$ production.  The $X$ is exchanged either (not both) in the $s-$ or $t-$/$u-$channel. $q$ may be $u$ or $d$.\label{Xplot1}}
\end{minipage}\hfill
\begin{minipage}{0.3\textwidth}
   \includegraphics[scale =0.3] {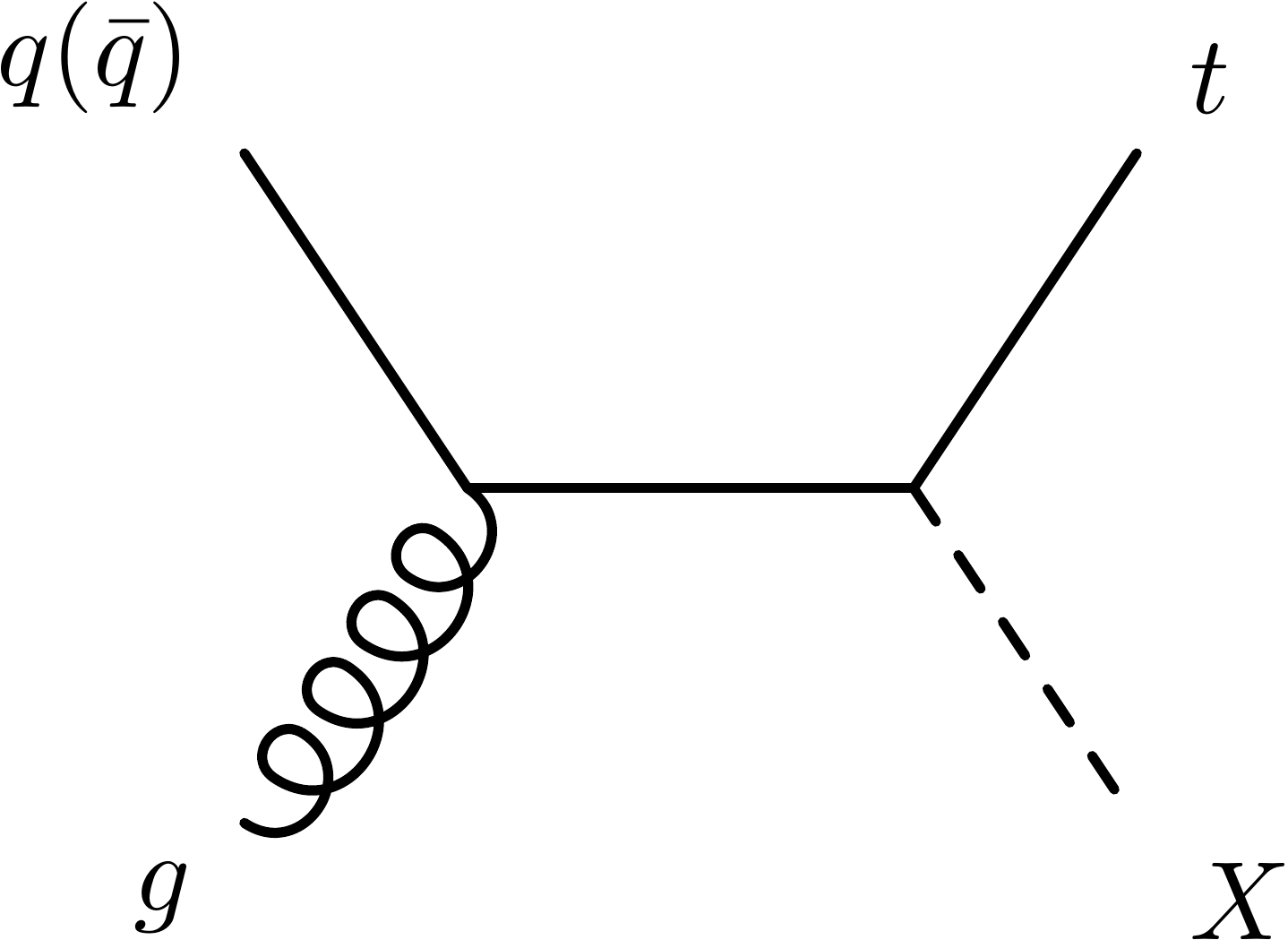}
\caption{For a $t$- or $u$-channel mediator $X$, direct production of $tX$ (followed by $X\to \bar t+q$ or $\bar q$) is always possible.\label{Xplot2}}
\end{minipage}
\end{figure}

No significant problems with the SM calculation or the experimental
measurements of the anomalously large asymmetry have been found.
Meanwhile, a variety of models have been proposed to explain it.
Most of these produce the asymmetry through the exchange of a new
particle, either an $s$-channel mediator with axial couplings to
both top and light quarks
\cite{Sehgal:1987wi,Bagger:1987fz,Djouadi:2009nb,Ferrario:2009bz,Frampton:2009rk,
Chivukula:2010fk,Bauer:2010iq,Chen:2010hm,Alvarez:2010js,Delaunay:2011vv,
Bai:2011ed,Barreto:2011au,Foot:2011xu,Zerwekh:2011wf,Haisch:2011up,Barcelo:2011fw,
Barcelo:2011vk,Gabrielli:2011jf,Tavares:2011zg,Alvarez:2011hi,AguilarSaavedra:2011ci},
or a $t$-channel (or $u$-channel) mediator
\cite{Jung:2009jz,Cheung:2009ch,Shu:2009xf,Arhrib:2009hu,Dorsner:2009mq,
Barger:2010mw,Xiao:2010hm,Cheung:2011qa,Shelton:2011hq,Berger:2011ua,Grinstein:2011yv,
Patel:2011eh,Craig:2011an,Ligeti:2011vt,Jung:2011zv,Buckley:2011vc,
Nelson:2011us,Duraisamy:2011pt,Cao:2011hr,Stone:2011dn} with
flavor-violating couplings that convert a light quark or antiquark
to a top quark. Both processes are illustrated in Fig.~\ref{Xplot1}.
In \cite{Jung:2009pi,Cao:2009uz,Cao:2010zb,Jung:2010yn,
Jung:2010ri,Choudhury:2010cd,Delaunay:2011gv,Gresham:2011pa,Shu:2011au,Gresham:2011fx,Westhoff:2011tq,Berger:2011},
comparisons between different models are carried out, and study of
those models or measurements in the LHC context can be found in
\cite{Gresham:2011dg,Blum:2011up,AguilarSaavedra:2011vw,Hewett:2011wz,
AguilarSaavedra:2011hz,AguilarSaavedra:2011ug,Jung:2011id,
Kahawala:2011sm,Berger:2011xk,Grinstein:2011dz,Berger:2011sv,Falkowski:2011zr}.

Charge asymmetries at the LHC are known to be powerful tools
for searching for and studying new physics, and recently this has
been put to use in the context of models for the $t\bar t$ asymmetry.  In
\cite{Craig:2011an} a large overall charge asymmetry was used to
argue the Shelton-Zurek model \cite{Shelton:2011hq} was most likely
excluded; a similar method was then applied for a different model in
\cite{Rajaraman:2011rw}. Here, we focus on models with $t$- or $u$-channel mediators, which, as we
will see, often generate large charge asymmetries in $t\bar tj$ (top
plus antitop plus a jet) at the LHC. These asymmetries, a
smoking gun of this type of model, will be crucial for a
convincing discovery or exclusion of this class of models.
Note these asymmetries are {\it not} directly related
to the Tevatron forward-backward asymmetry in $t\bar t$ events,
which translate at the LHC into the differential charge asymmetry in $t$ production mentioned
above.  The asymmetry in $t\bar t j$ that we study here stems from a completely different source; see below.

Any of the models with a $t$- or $u$-channel mediator has a coupling
between a light quark or antiquark, a top quark, and a new particle
$X$, as in Fig.~\ref{Xtchannel}.  It follows that the $X$ can be
produced from an off-shell quark or antiquark in association with a
$t$ or $\bar t$, as shown in Fig.~\ref{Xplot2}. Consequently, as has
been pointed out by many authors
\cite{Shu:2009xf,Dorsner:2009mq,Cheung:2011qa,Gresham:2011dg,Shu:2011au,Berger:2011xk},
it is important at the LHC to look for the process $pp\to Xt$ (and the conjugate process $pp\rightarrow\bar X\bar t$), where
$X$ in turn decays to $\bar t$ plus a jet.  A straightforward search
for a $t$+jet resonance can be carried out, though it suffers from
the poor resolution for reconstructing the resonance, large
intrinsic backgrounds whose shape may peak near the resonance mass,
and combinatoric backgrounds in the reconstruction. Alternatively,
one could attempt a cut-and-count experiment; with appropriate cuts
one can obtain samples in which the $X$ production contributes a
statistically significant excess to the $t\bar tj$ rate. But the
$t\bar tj$ background is not simple to model or measure, and
systematic errors may be problematic.

Fortunately, the process shown in Fig.~\ref{Xplot2} has a large
charge asymmetry. The difference between quark and antiquark pdfs
assures that the rate for $X$ production is different from that of
$\bar X$ production.  (If $X$ is self-conjugate, same-sign top-quark
production results, and is readily excluded
\cite{ATLASSameDiTop,Chatrchyan:2011dk}; we therefore assume that $\bar X\neq X$.) 
Our approach in this work will be to suggest something a
bit more sophisticated than a simple resonance search, using the
charge asymmetries of these models to reduce systematic errors at a
limited price in statistics.  We will also propose other
charge-asymmetric variables that can serve as a cross-check.  As a
by-product, should any discovery occur, the asymmetry itself can
serve as a diagnostic to distinguish certain classes of models from
one other.

\section{Benchmark Models}

As our benchmark model, we take a typical model with a $t$-channel mediator,
a colorless charged spin-one particle which we call a $W'$.  We will assume the $W'$ couples a
right-handed $d$ quark to a $t$ quark.  While a theory with only these
couplings would be inconsistent, we will assume this coupling
generates the largest observable effects.  One may say that we choose
a ``simplified model'', or ``model fragment'', in which this coupling
is the only one that plays an experimentally relevant role.  We will
see this point is not generally essential.\footnote{Attempts to make
  consistent models with a $W'$ include \cite{Barger:2011ih}.  There are also
  attempts to include the coupling of a $W'$ with a $u$ and $b$
  quark \cite{Shelton:2011hq}, but such couplings lead to a large charge
  asymmetry in single top production \cite{Craig:2011an}, now excluded by LHC
  data \cite{ATLASSingleTop,Chatrchyan:2011vp}.}  The Lagrangian we take for our simplified model
is simply
\begin{equation}
\mathcal {L}=-g_R W_{\mu}^{'+} \bar{t}\gamma^{\mu} P_R d +h.c.
\end{equation}
where $P_R=(1+\gamma^5)/2$.

\begin{figure}[t]
\centering
\subfigure[$W'^-$ production]{
   \includegraphics[scale =0.4] {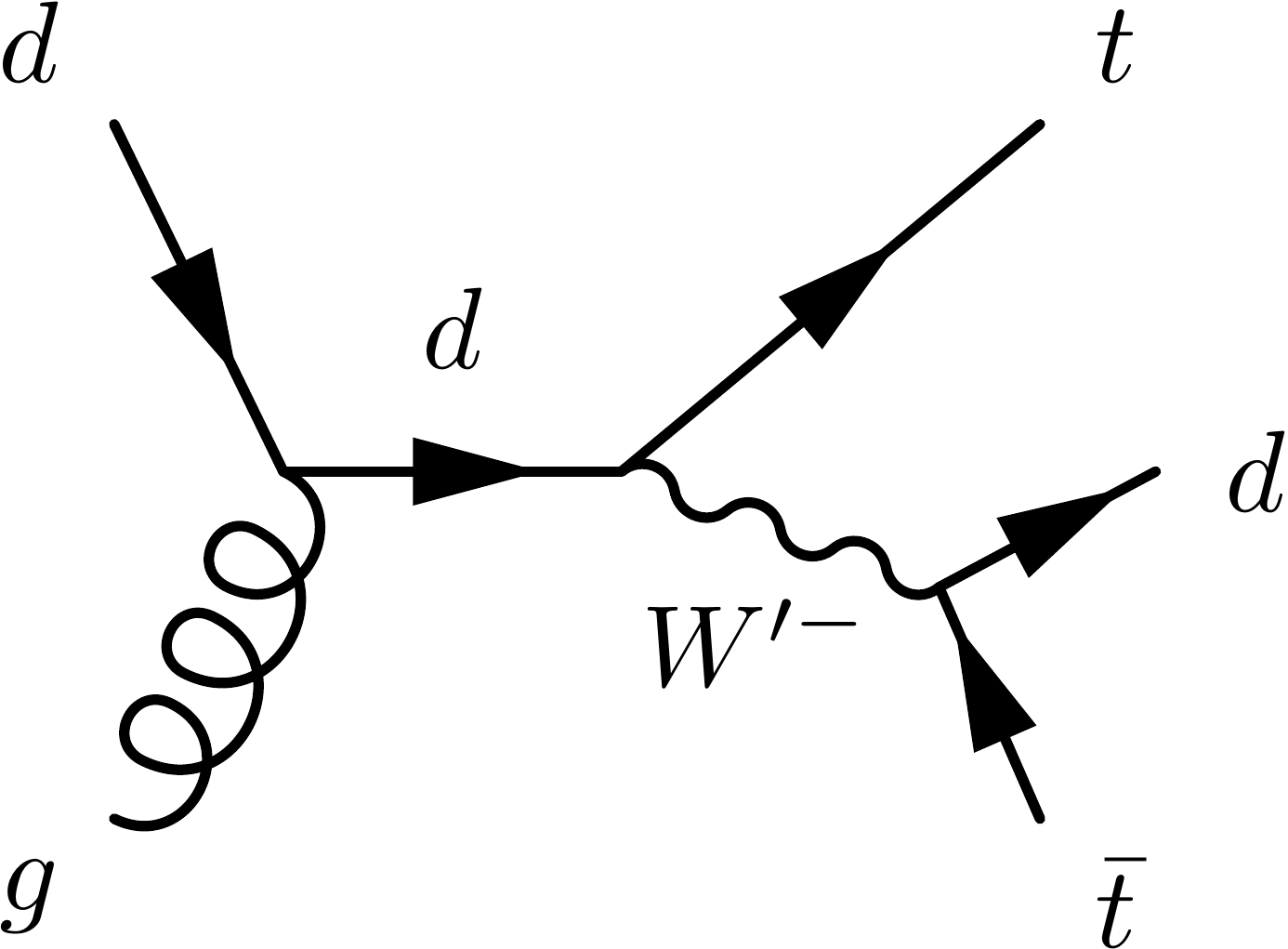}
   \label{wpfeynmin}
 }\hfill
 \subfigure[$W'^+$ production]{
   \includegraphics[scale =0.4] {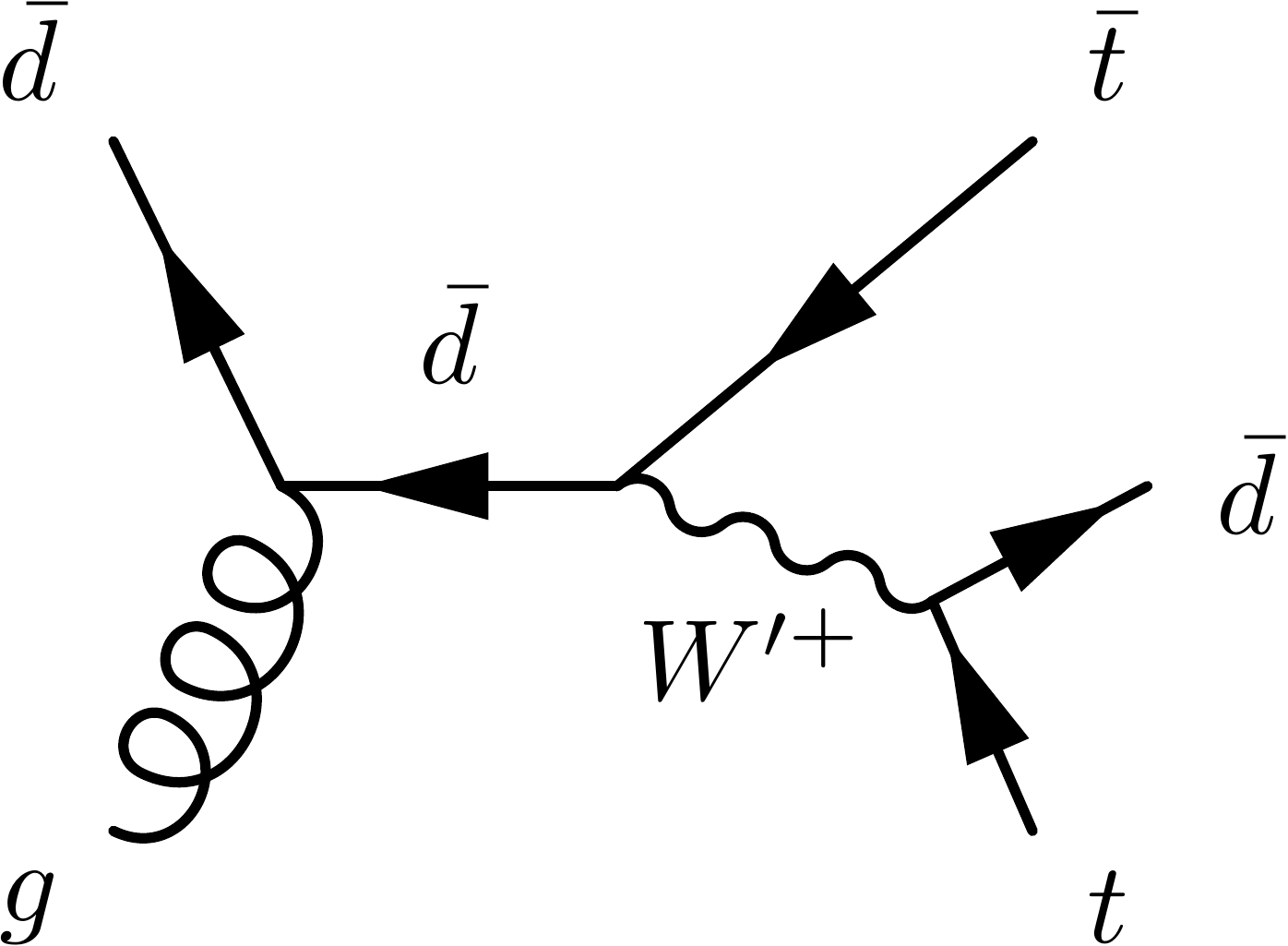}
   \label{wpfeynplus}
 }
\caption{Dominant production mode for the $W'$. The cross-section for
$W'^-$ is much larger than for $W'^+$.\label{schannel}}
\end{figure}

\begin{figure}[t]
\centering
\begin{minipage}{0.3\textwidth}
   \includegraphics[scale =0.3] {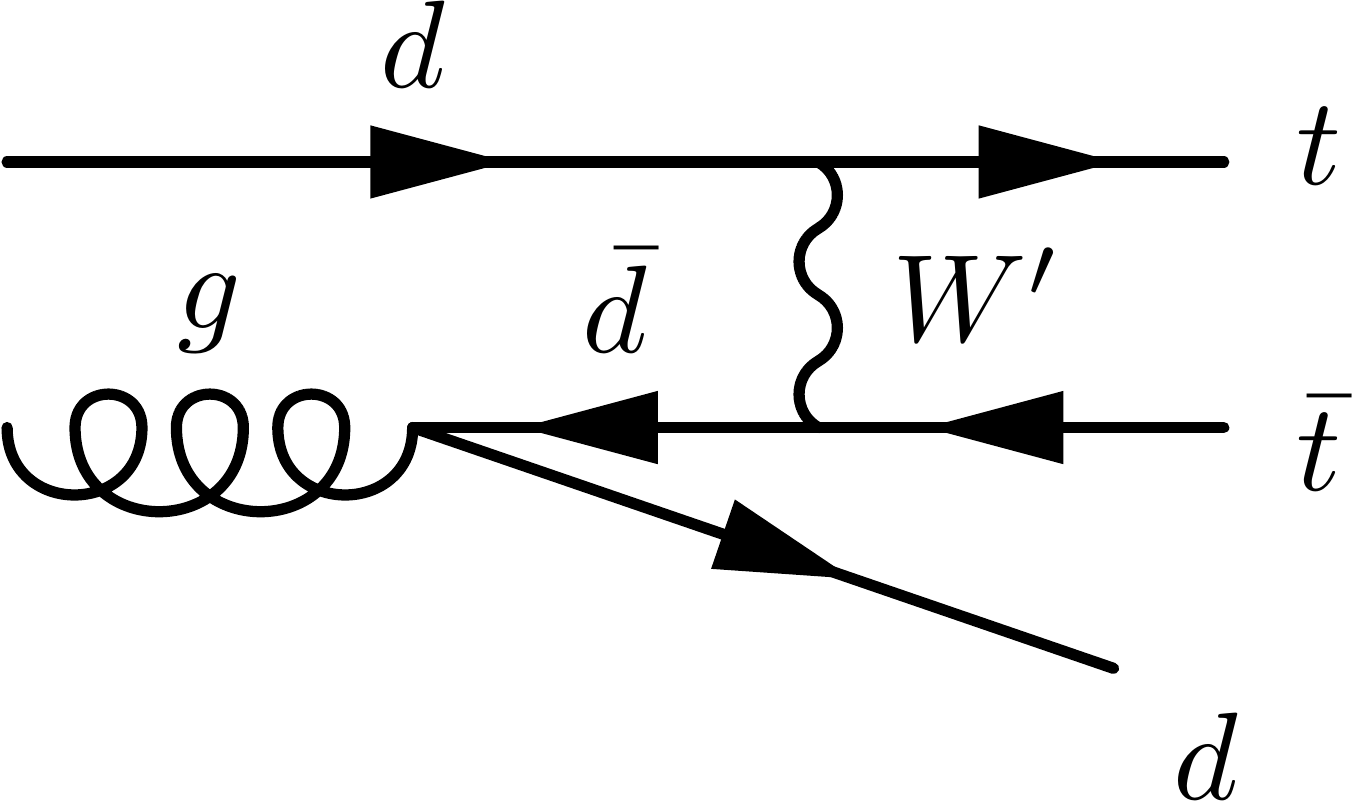}
\end{minipage}\hfill
\begin{minipage}{0.3\textwidth}
   \includegraphics[scale =0.3] {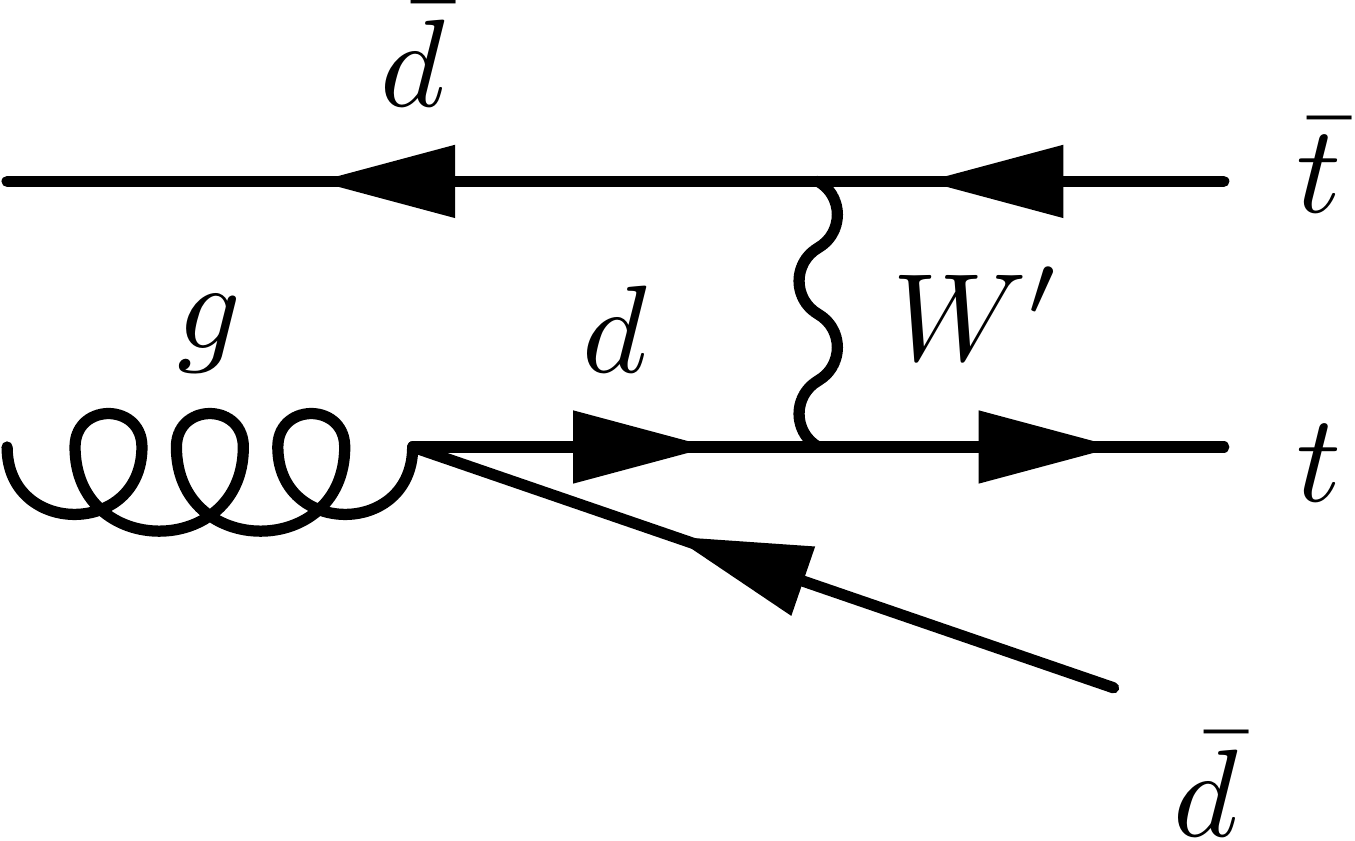}
\end{minipage}
 \hfill
\begin{minipage}{0.3\textwidth}
   \includegraphics[scale =0.3] {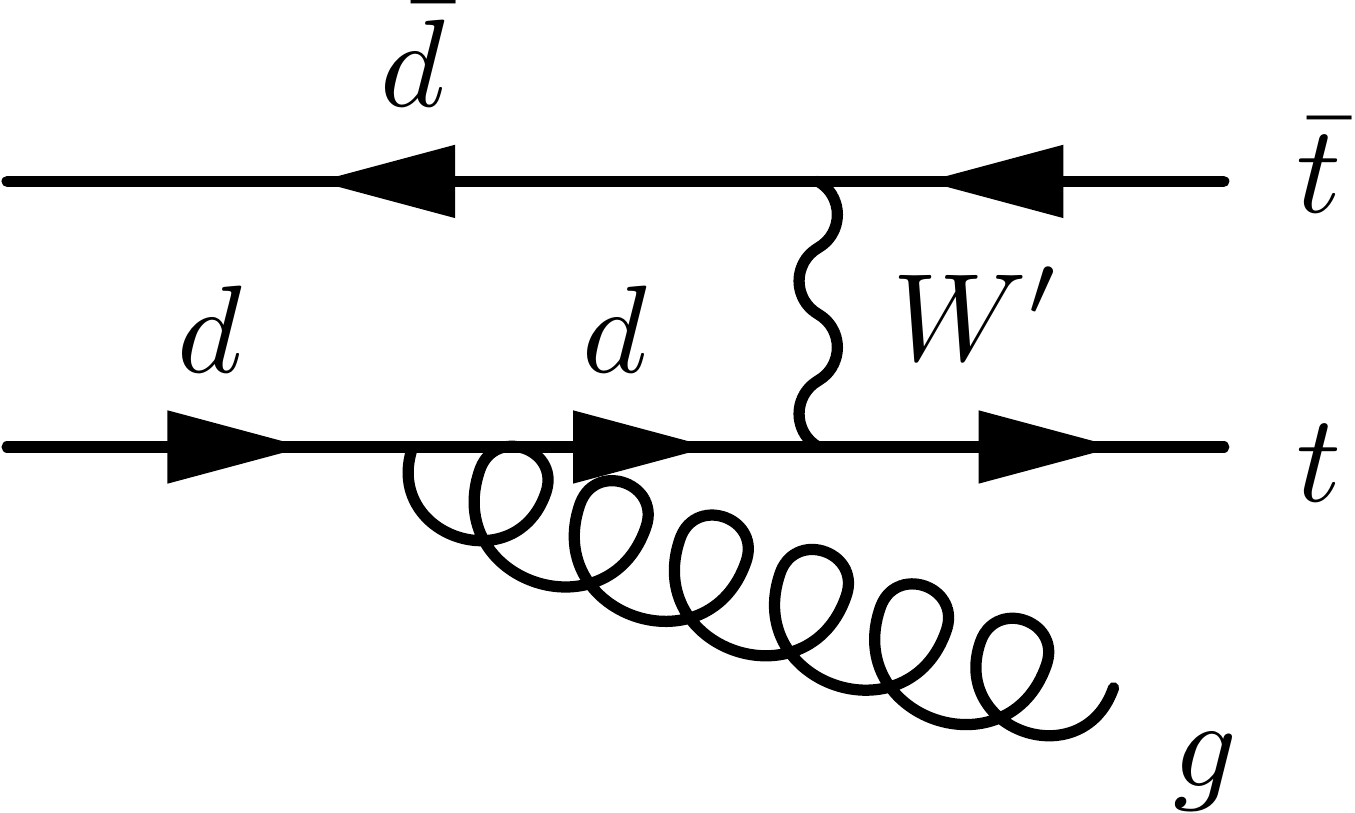}
\end{minipage}
\caption{Characteristic examples of diagrams that
contribute to $t\bar t  j$  production involving the $W'$ in $t$-channel
exchange.\label{tchannel}}
\end{figure}

We are interested in the process in which the $W'$ contributes to a
$t\bar t j$ final state. One contribution comes from $d g\to t W'^-$
and its conjugate $\bar d g\to \bar t W'^+$, following which the
$W'^-$ decays to $\bar t d$ and the $W'^+$ decays to $t \bar d$.  We
will refer to this as ``$s$-channel production'' (see
Fig.~\ref{schannel}).  The $W'$ also contributes to $d g\to t\bar t
d$, and similar processes, through $t$-channel exchange (see
Fig.~\ref{tchannel}).

The cornerstone of our analysis is the observation that \emph{in the $s$-channel process, the
negatively charged $W'$ is produced more abundantly than the
positively charged $W'$}, because the negative $W'$ can be produced from a
valence quark, while a positive $W'$ requires a sea antiquark in the initial
state. (See Fig.~\ref{schannel}.)

The processes in Figs.~\ref{schannel} and \ref{tchannel}
 can in principle have
non-trivial interference with the Standard Model background --- a
point which considerably complicates background simulation.
But we have found that interference is not numerically
important for certain observables, at least with
current and near-term integrated luminosities. {\it  All results in this
paper therefore ignore interference; however, with larger data sets, or when studying other models and/or using other variables, one must confirm on a case-by-case basis that this approximation is sufficiently accurate for the analysis at hand.}

In \cite{Cheung:2011qa}, the authors studied this model and fitted
it to the $t\bar{t}$ asymmetry and total cross-section in CDF. (This
was done prior to the DZero result that shows a smaller asymmetry
with less energy dependence.) Based on this work, we will take six
benchmark points shown in Table \ref{crosssections}, with three
values of the $W'$ mass and two values of $g_R$ for each mass, a
larger value that would reproduce the CDF measurement and a value
$\sqrt 2$ smaller that would give a Tevatron asymmetry (and also an
$W'$ width and $tW'$ production rate) of about half the size. The
cross-sections at these benchmark points (including all the
processes shown in Figs.~\ref{schannel} and \ref{tchannel}) are also
given in Table \ref{crosssections}.

\begin{table}
\renewcommand{\arraystretch}{1.2}
\begin{tabular}{|c|c|c|}
\hline
Mass (GeV)&$\:\:g_R\:\:$&cross-section (pb)\\
\hline\hline
400  & 1.5 &32.2\\
400  & $\frac{1.5}{\sqrt{2}}$ &12.9 \\
600  & 2&18.2\\
600 & $\sqrt{2}$&6.3 \\
800 & 2 &6.5 \\
800 & $\sqrt{2}$ &2.1 \\[0.7ex]
\hline
\end{tabular}
\caption{7 TeV LHC tree-level cross-sections for the processes shown in
Figs.~\ref{schannel} and \ref{tchannel}, for the various benchmark
points. No K-factor is included in these numbers, but we do apply one later in our analysis; for a discussion of the simulations and the K-factor, we refer the reader to Sec.~\ref{Seceventselection}. \label{crosssections}}
\end{table}

The $W'$ also contributes to $t\bar t$ production through
$t$-channel exchange, and thus to the differential charge asymmetry
in $t$ rapidity at the LHC (not to be confused with the asymmetries
in $t\bar tj$ that are the subject of this paper.)  ATLAS and CMS
measurements of this quantity (with respectively 0.7 and 1.1 $\mathrm{fb}^{-1}$ of data)
\cite{ATLASChargeAsymm,CMSChargeAsymm} may somewhat disfavor the
benchmark points with the larger values of $g_R$, which (at parton-level, not accounting for $t$ reconstruction efficiencies) give a
differential charge asymmetry in the 8--9\% range. But the situation
is ambiguous, since event mis-reconstruction and detector resolution
produce a large dilution factor, which may make this charge
asymmetry consistent with the current measurements. Our benchmarks
with larger couplings thus probably represent the outer edge of what might
still be allowed by the data.  By considering also an
intermediate coupling that still could explain the Tevatron $t\bar
t$ asymmetry, we cover most of the interesting territory, and permit
the reader to interpolate to other values of the couplings.

\section{A mass variable\label{Secvar}}

Among the charge-asymmetric observables discussed in this paper, we
will devote most of our attention to one motivated by the resonance
structure of the $W'$, which we will refer to as the mass variable
$M_{j1bW}$ in later content. This variable is applicable universally
to a wide range of $W'$ masses and couplings, and to most other
models with $tX$ production.  We discuss this mass variable in great
detail in this section.  In Sec.~\ref{Secangle}, we will discuss
the azimuthal angle between the hardest jet and the lepton (which we
refer to as the ``angle variable''.) A third class of potentially
useful variables (``$P_T$ variables''), including the $P_T$
difference between the hadronic and the leptonic top quarks or
$W$-bosons, is briefly discussed in Appendix~\ref{AppPtdiff}.

We will consider only the
semi-leptonic $t\bar tj$ events (where one top decays hadronically and the other
leptonically), resulting in a final state of 5 jets, a lepton and missing
energy.  All-hadronic decays are not useful for a charge asymmetry, as $t$
and $\bar t$ cannot be distinguished in this case, while the fully leptonic decay, though probably useful, has a low
branching fraction.

Since it is the $s$-channel process in Fig.~\ref{schannel}, where the $W'$ appears as a resonance, that is charge-asymmetric, we will focus our attention there.
In our later analysis we will impose an $S_T$ cut\footnote{For our definition of $S_T$, see equation (\ref{Stdef}) in Sec.~\ref{Seceventselection}.} to improve the signal-to-background ratio.
If we put that cut at $700$ GeV, the fraction of negatively charged
$W'$s for the 400, 600 and 800 GeV $W'$ is $0.84$, $0.87$ and $0.86$
respectively.  Such an enormous charge asymmetry in production can be
put to good use.

Note, however, that since every event (following the $W'$ decay) has a
$t$ and a $\bar t$, either of which may produce the lepton, \emph{the
  total numbers of events with positively and negatively charged leptons are
  expected to be roughly equal}, up to edge effects produced by cuts
and detector acceptance. But since negative $W'$s are produced more
abundantly, \emph{a negatively charged lepton is more likely to come
  from the $W'$ decay, while positive leptons tend to originate from
  the decay of the spectator top quark or antiquark.}  Kinematic
features, such as the invariant mass and transverse mass of various
final-state objects, differ for events with negatively and positively charged
leptons.  For instance, a simple bump hunt aimed at reconstructing the
$W'$ resonance would find a much larger bump in negatively charged
leptons than in positively charged ones.  Here, we will consider the
$W'$ reconstructed mass distribution more completely, noting that
\emph{the signal remains asymmetric even away from the $W'$ mass
  bump}, since the total asymmetry must integrate to (almost) zero.

Another useful kinematical feature is that the hardest jet in $tW'\to t\bar td$ production commonly originates
from the $d$-quark, because of the large energy released in the $W'$ decay and the dissipation
of the top quarks' energies into their three daughters. At leading order and
at parton-level, and with an $S_T$ cut of 700 GeV,
the fraction of
events where the
hardest parton is the $d$-quark (or antiquark) from the $W'$ is 0.71, 0.82 and 0.82
for a $W'$ of mass 400, 600 and 800 GeV respectively. (Note neither ISR/FSR, hadronization, nor jet reconstruction are accounted
for in these numbers, which are for illustration only.)
We have designed our variables to maximally exploit these two
kinematic features.

One conceptually simple approach to seeking the $W'$ would involve
fully reconstructing the $t$ and $\bar t$ in each event, and searching
for a resonance in either $tj$ or $\bar tj$. This has been discussed
in \cite{Shu:2009xf,Dorsner:2009mq,Cheung:2011qa,Gresham:2011dg,Shu:2011au,Berger:2011xk}. The
challenge is that the combinatoric background is large and hard
to model, and often peaks in a region not far from the
resonance. Charge-asymmetries are useful here, because the
positive-charge lepton events are dominated by the combinatoric
background, while the negative-charge lepton events have similar
combinatorics but a much larger resonance.  Comparison of the two
samples would allow for the elimination of a significant amount of systematic error.

However, full event reconstruction in events with five jets will have low efficiency, and moreover we are neither confident in our ability to model it nor certain it is the most effective method.  Here we will instead focus on variables that require only partial event reconstruction. Of course the experimental groups should explore whether full event reconstruction is preferable to the methods we attempt here.

\begin{figure}[t]
\centering \subfigure[$M_{j1bW}$ (parton-level) for  signal only, shown for positive and negative lepton charge.]{
   \includegraphics[width =0.45\textwidth] {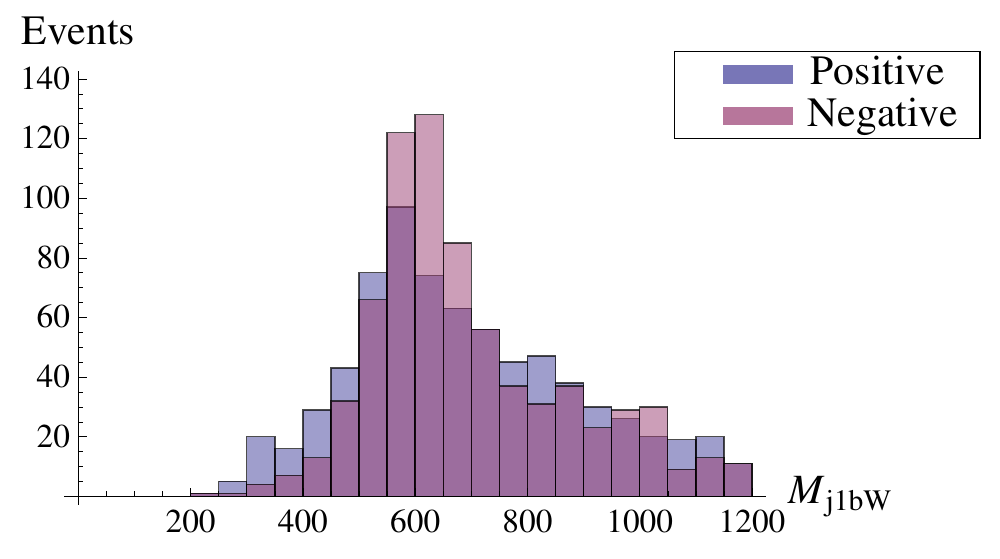}
   \label{wplhehist}
 }\hfill
 \subfigure[Bin-by-bin signal-only charge asymmetry]{
   \includegraphics[width =0.45\textwidth] {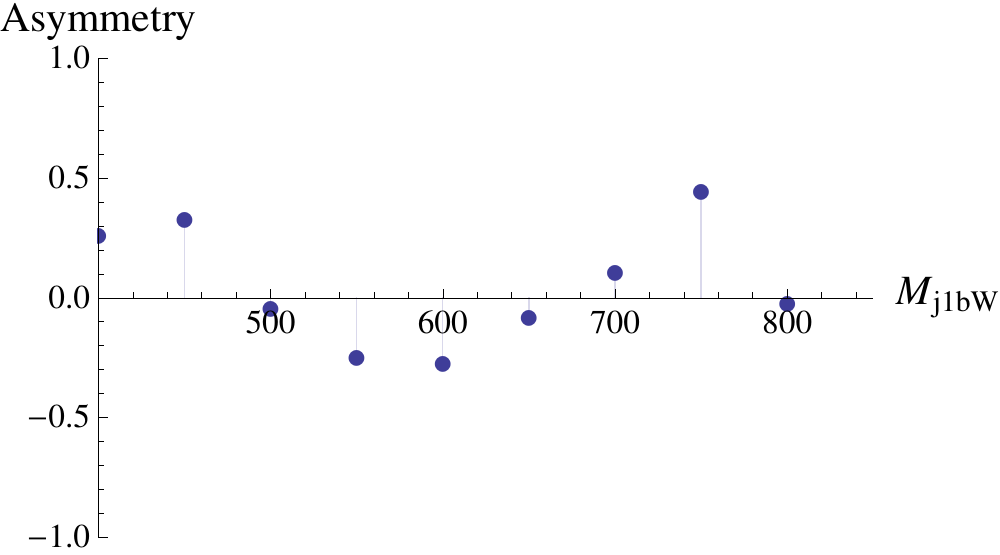}
   \label{wplheas}
 }
\subfigure[$M_{j1bW}$ (parton-level) for signal plus background, shown for positive and negative lepton charge.]{
   \includegraphics[width =0.45\textwidth] {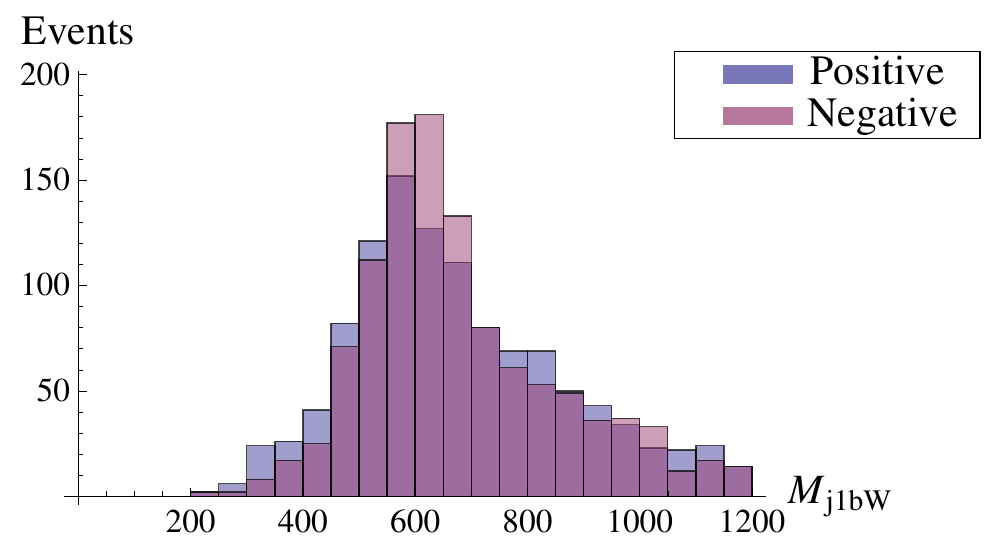}
   \label{SBlhehist}
 }\hfill
 \subfigure[Bin-by-bin signal plus background charge asymmetry]{
   \includegraphics[width =0.45\textwidth] {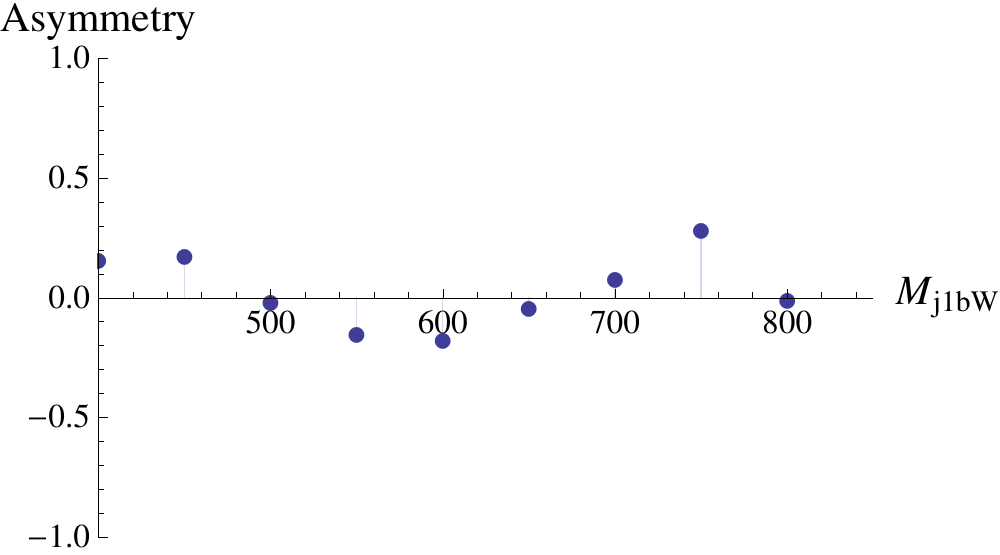}
   \label{SBlhehist}
 }
\caption{Parton-level charge asymmetry in the $M_{j1bW}$ variable for
a 600 GeV $W'$ with $g_R=2$ and an $ S_T$ cut at 700 GeV. The leptonic
$W$ boson was reconstructed from its decay products, $j_1$ was
taken to be the hardest non-$b$ parton.  ISR/FSR and $b$-quark
selection effects were not accounted for here. The sample
corresponds to 1.5 fb$^{-1}$. \label{wplhe}}
\end{figure}

 We will focus on the mass variable $M_{j1bW}$: the invariant mass of the hardest jet in the event, a $b$-tagged jet (chosen as described below), and a $W$-candidate reconstructed from the observed lepton and the missing transverse momentum (MET).\footnote{We solve for the neutrino four-momentum in the usual way. Complex solutions are discarded for simplicity. When two real solutions exist, the most central $W$ candidate is selected.}  It involves only a partial reconstruction of the event to
form a candidate for the $W'$, assuming it has decayed to a lepton.\footnote{Were one to fully reconstruct the $t\bar tj$ events, one could also study  the invariant mass of the {\it hadronically}-decaying top and
the  hardest jet, which will also differ for positive- and negative-charge lepton events. 
We neglect this variable here because the reconstruction of the hadronic top has low efficiency, but we encourage our experimental colleagues to consider if they can increase their sensitivity by including it.}
In signal events where the hardest jet in the event is a $d$ (or $\bar d$) from the $W'$ decay, and the $\bar t$ (or $t$) from the $W'$ produces a lepton $\ell$, $M_{j1bW}$ often reconstructs the $W'$ resonance. The events with an $\ell^-$ typically exhibit a resonance at the $W'$ mass, while
those with an $\ell^+$, in which the $W'$ is most often not reconstructed
correctly, have a smoother distribution. This effect, and the resulting charge asymmetry --- with a negative asymmetry near the $W'$ mass and positive asymmetry elsewhere --- are
shown for $m_{W'}=600$ GeV in Fig.~\ref{wplhe}.  Both the asymmetric $s$-channel
and the almost symmetric $t$-channel are included in what we call ``signal.''

In constructing $M_{j1bW}$, we reduce the combinatorial background by rejecting
$b$-jets that are inconsistent with forming  a top quark with the
lepton and the MET ({$M_{bl}<155$ GeV and $M^T_{bl\nu}<175$ GeV.) When
multiple $b$-jets satisfy these criteria, we select the $b$-quark for which the quantity $|M_{bl}-155\:\mathrm{GeV}|+|M^T_{bl\nu}-175\:\mathrm{GeV}|$ is smallest.  The combined
efficiency of the $W$ reconstruction and the $b$ selection is about $45\%$.

Meanwhile, we will give evidence in Sec.~\ref{SMas} that the SM
background to this process shows no charge asymmetry in this
variable, to a sufficiently good approximation.  It is crucial for the
use of this variable that this is true.

There are other invariant-mass and transverse-mass variables that have
their merits. Some require no event reconstruction, including the
invariant mass of the hardest jet and the lepton ($M_{j1l}$) and the
invariant mass of the hardest jet, a $b$-tagged jet and the lepton
($M_{j1bl}$). For quantities that include the MET in the event, one
could consider the transverse mass of two or more objects.  (See
also the footnote above concerning the hadronically decaying top in
fully reconstructed events.) These variables and their charge asymmetries are strongly correlated, but one might still obtain additional sensitivity by combining them.  But here,
for simplicity, having found that the most sensitive variable on its
own is $M_{j1bW}$, we will focus on it exclusively below.

\section{Event selection and processing\label{Seceventselection}}

We mentioned earlier that the $t\bar tj$ background and the $W'$ signal do interfere
with each other.  However we have
explicitly checked that interference effects do not alter
the differential asymmetry in the $M_{j1bW}$ mass variable by a significant
amount (given currently
expected statistical uncertainties).  The effect on the total
number of events is also small.  Thus it is relatively safe for us ---
and for the early searches at the LHC --- to neglect interference in the study
of the mass variable, at least for the $W'$ model.  (We have not studied whether this is true for
all similar models with $tX$ production.)  At some point, higher-precision study with much
larger data samples ($\gg$ 10 fb$^{-1}$) may require the full set of interfering diagrams, and
a special-purpose background-plus-signal simulation.  Here we simulate background and
signal independently.

On the other hand,
$t$-channel $W'$ exchange (Fig.~\ref{tchannel}) makes
an important contribution to the cross-section and should always be included when generating
the signal sample. (This is not uniformly the case in the literature.)
For the variables we are studying, the $t$-channel process does not contribute much to the asymmetry,
and effectively acts as an additional background.

A background sample and the signal samples for our benchmark points
were generated with Madgraph 4.4.32 \cite{Alwall:2007st} and
showered with PYTHIA 6.4.22 \cite{Sjostrand:2006za}.  We performed a fast detector simulation
with DELPHES 1.9 \cite{delphes}. (For our parton-level studies the decays of
the top and the antitop were simulated with BRIDGE 2.24
\cite{Meade:2007js}). We used the anti-$k_T$ jet-clustering
algorithm (with $R=0.5$) to reconstruct jets. The isolation of
leptons and jets is described in Appendix~\ref{AppIsolation}. The
$b$-tagging was modeled after the SV050 tagger of the ATLAS
collaboration \cite{btag}. We account for the rising
$P_T$-dependence of the $b$-tagging efficiency, which reaches up to
$60\%$ in the kinematic regime of interest. The dependence of the
$b$ tagging efficiency on the pseudo-rapidity is assumed to be
negligible within the $\eta$ reach of the tracker ($|\eta|<2.4$),
with the tagging rate taken to be zero outside the tracker. The
$c$-tag efficiency was assumed a factor of 5 smaller and the mistag
rate is taken to be $1\%$.  We do not account for the falloff in
efficiency and the rise in mistag rates at higher $P_T$, since
measurements of these effects are not publicly available; our
tagging might therefore be optimistic, though the issue affects both
signal and background efficiency.

We impose the following criteria for our event
selection:\footnote{Our cuts may be optimistic in the
rapidly changing
LHC environment.
Raising the jet $P_T$ cut to 40
GeV results in a loss of sensitivity of order 10--20\%.
If one restricts jets to those with $|\eta|<2.5$, signal
is reduced by about 10\%, and background by about 15\%.
An increase in the electron
$P_T$ cut to 45 GeV reduces signal by 20--25\% and background
by about 30\%.}
\begin{itemize}
\item At least 5 jets with $P_{T}^{jet}>30$ GeV and $|\eta|<5$
\item At least one of these jets is $b$-tagged
\item One isolated lepton ($e^\pm$ or $\mu^\pm$) with $P_{T}^\ell>30$ GeV and
$|\eta|<2.5$
\item MET $>30$ GeV.
\end{itemize}
where $\eta$ stands for pseudo-rapidity as usual.
We also impose a cut on $S_T$, which is defined as
\begin{equation}
S_T=\sum P_{T}^{jet}+P_{T}^{\ell}+\mathrm{MET} \label{Stdef}
\end{equation}
 where the sum runs over all the jets with  $P_{T}^{jet}>30$ GeV. The $S_T$ cut will be at a high enough
scale (typically 600-800 GeV) that our events will pass the trigger with high efficiency.

The SM background simulation requires a matched sample for
\begin{align*}
p+p&\rightarrow t+\bar{t}\\
p+p&\rightarrow t+\bar{t}+j
\end{align*}
where we use the MLM scheme \cite{Caravaglios:1998yr}, with QCUT$=30$
and xqcut$=20$. The renormalization and factorization scales are set to $m_T$, where
$m_T^2$ is the  the geometric mean of $m_t^2+p_T^2$ for the top and
antitop.

One might wonder whether it is necessary to include $pp\to t\bar
tjj$ as well. But we are requiring 5 hard jets, and the mass and
angle variables we will study are not sensitive to soft radiated
jets, as they involve the hardest jet and a $b$-tagged jet. It is
sufficient, therefore, for us to truncate our matched sample with
one jet, and allow PYTHIA to generate any additional radiation. In
total, we generated 3 million background events before matching.
After matching, we find an inclusive $t\bar{t}$ LO cross-section of
about 90 pb, so we include a K-factor of 1.7 to match with the
NLO+NNLL QCD calculation \cite{Ahrens:2011mw,QCDNLO}. The number of
events we generated for background corresponds to about 14
$\mathrm{fb}^{-1}$, large enough to provide smooth distributions for the
variables we study.

There are a number of SM processes whose {\it total cross-sections} for producing
a lepton are intrinsically charge-asymmetric.  These include single-top production and
$W$-plus-jets, for which an $\ell^+$ is more likely than an $\ell^-$.
However, these have small rates for 5 jets and a lepton, especially with a
$b$ tag required and with a hard $S_T$ cut.  Moreover, asymmetries
from any such process would be quite different from
the signal, being both structureless and everywhere positive.  We foresee no problem
with such backgrounds.

For each value of the $W'$ mass and coupling constant, we generated  a signal
sample with 750,000 events.  No matching was used; extra ISR/FSR jets were
generated by PYTHIA. These samples are large enough to suppress statistical
fluctuations when we later use them to study the expected shape and magnitude of
the asymmetry. In our studies, we have chosen to scale all LO signal cross-sections, for all six benchmark points, by a K-factor of 1.7, the same as for the $t\bar t$ background.\footnote{We note that the K-factor for the process $bg\rightarrow t W$ is in this range \cite{gbtW}, suggesting our choice is not unreasonable.} Note that this K-factor can always be absorbed in $g_R$, as long as the width of the $W'$ is smaller than the resolution.

\section{Analysis and results\label{Secstat}}

Although the parton-level charge asymmetries described in Sec.~\ref{Secvar} are
large, the experimentally observable asymmetries are  significantly diluted by the detector resolution and mis-reconstructions.
Fig.~\ref{wplhco} shows our estimate of the asymmetry structure
that can be obtained at the detector level; compare this with Fig.~\ref{wplhe}. Note, however, that
the basic structure of a negative asymmetry at the $W'$ peak, with a positive asymmetry to either side, remains intact.

\begin{figure}[h!]
\centering \subfigure[$M_{j1bW}$ for a 400 GeV $W'$, $g_R=1.5$]{
   \includegraphics[width =0.45\textwidth] {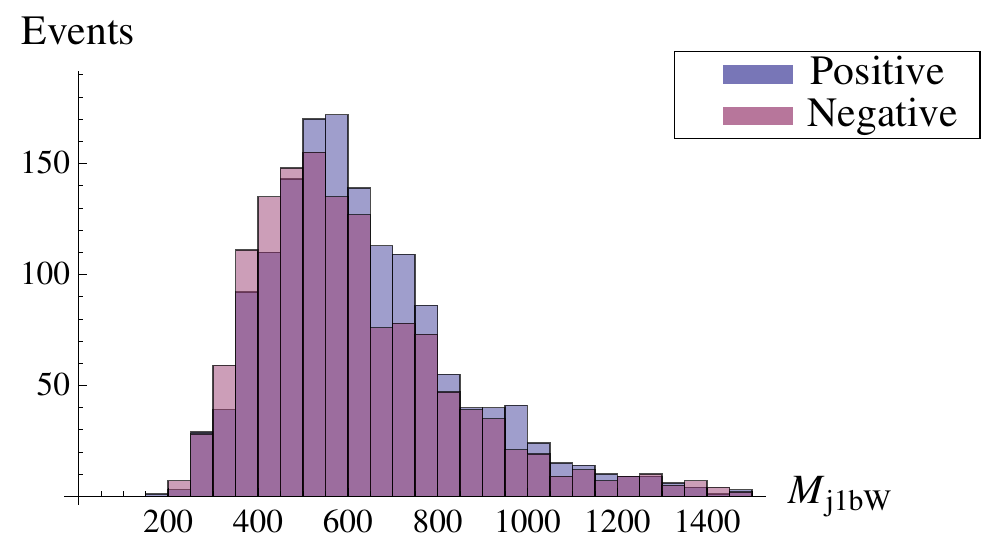}
   \label{wplhcohist}
 }\hfill
 \subfigure[Bin-by-bin asymmetry for a 400 GeV $W'$, $g_R=1.5$\label{lhcoas}]{
   \includegraphics[width =0.45\textwidth] {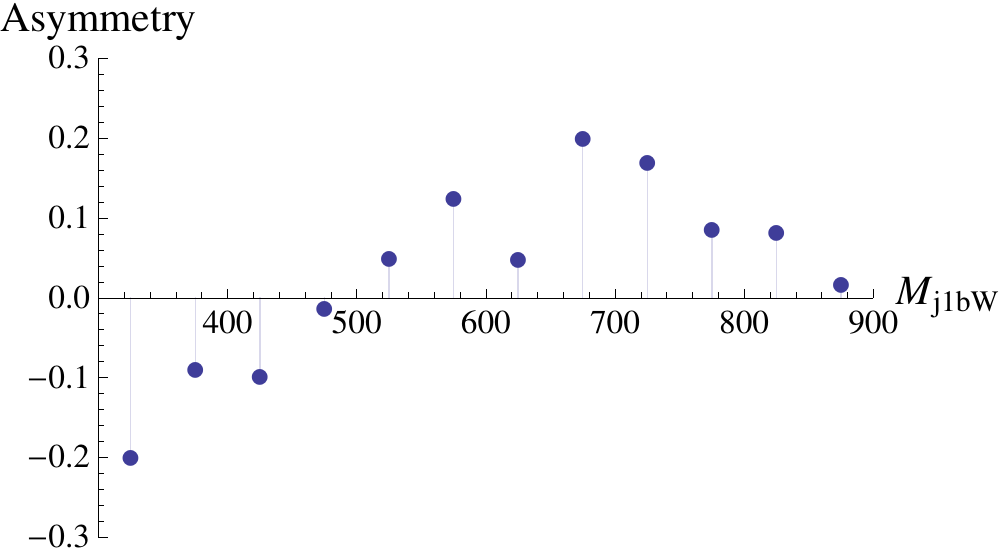}
   \label{wplhco}
 }
 \subfigure[$M_{j1bW}$ for a 600 GeV $W'$, $g_R=2$]{
   \includegraphics[width =0.45\textwidth] {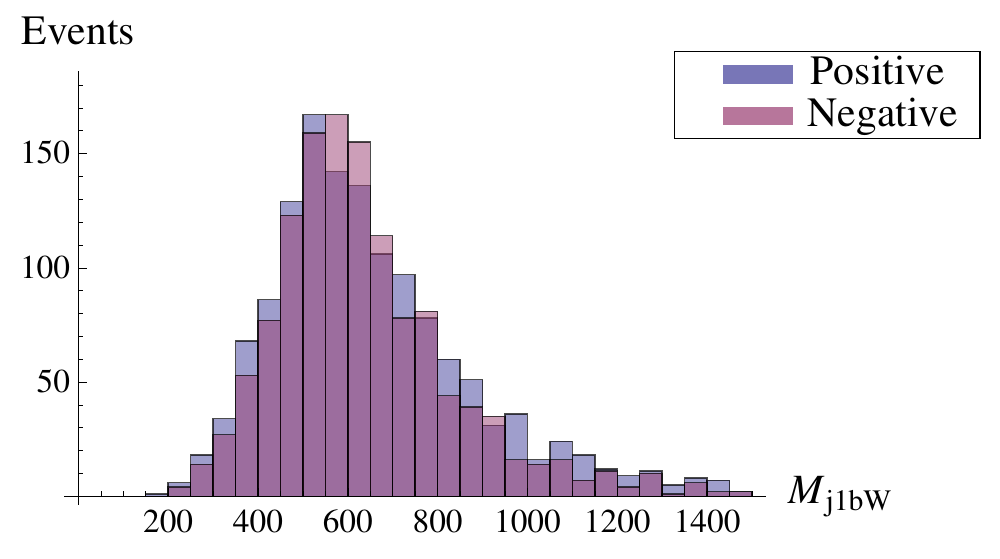}
   \label{wplhcohist}
 }\hfill
 \subfigure[Bin-by-bin asymmetry for a 600 GeV $W'$, $g_R=2$\label{lhcoas}]{
   \includegraphics[width =0.45\textwidth] {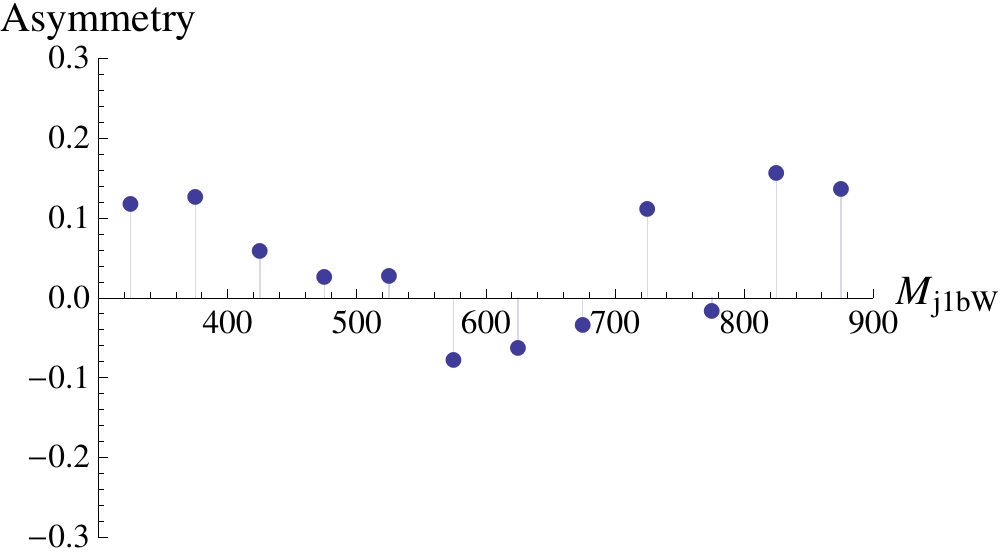}
   \label{wplhco}
   }
    \subfigure[$M_{j1bW}$ for an 800 GeV $W'$, $g_R=2$]{
   \includegraphics[width =0.45\textwidth] {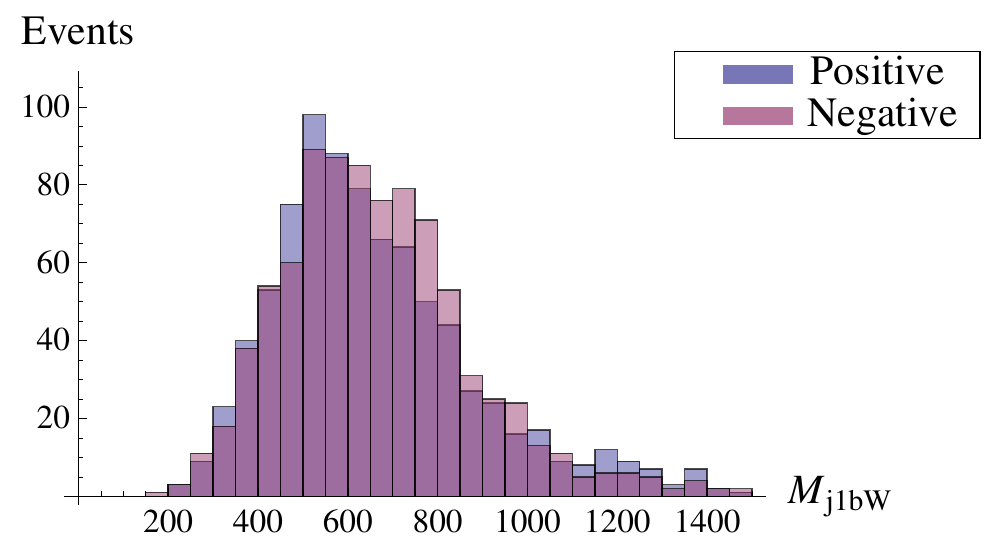}
   \label{wplhcohist}
 }\hfill
 \subfigure[Bin-by-bin asymmetry for an 800 GeV $W'$, $g_R=2$\label{lhcoas}]{
   \includegraphics[width =0.45\textwidth] {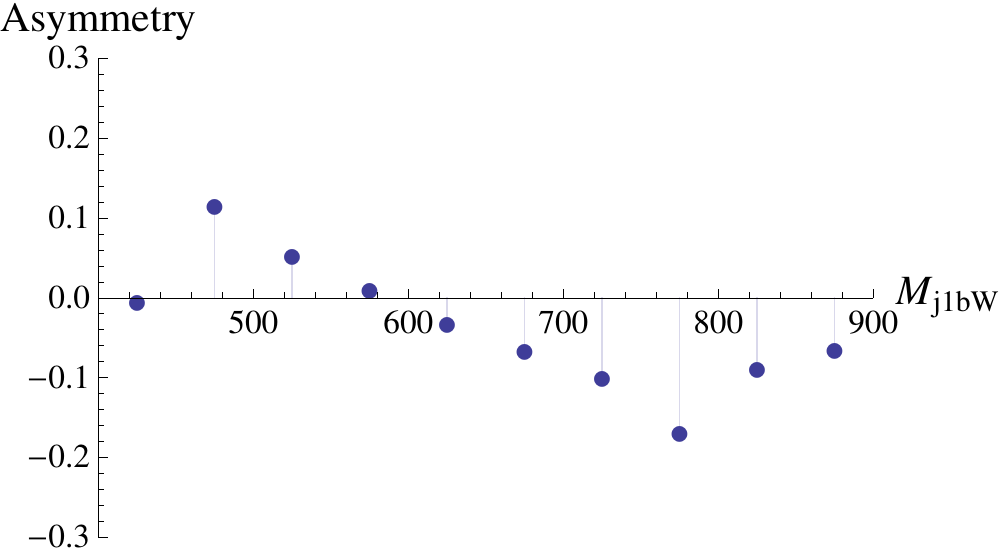}
 }
\caption{As in Fig.~\ref{wplhe}, but after accounting for detector effects, and with an $S_T$ cut of 700
GeV, for three different $W'$ masses.  All plots show
 signal plus background. The
samples correspond to an integrated luminosity of 5 $\mathrm{fb}^{-1}$.\label{wplhco}}
\end{figure}

As always, one needs to obtain a
prediction for both the Standard Model-only assumption (SM) and the
Standard Model plus new physics assumption (NP), and assign a degree
of belief to one or the other using a suitable statistical procedure,
given the observed data. We will argue below that the SM prediction for
the asymmetry in $M_{j1bW}$ is essentially zero, within the statistical
uncertainties of the measurement.  However, to predict the asymmetry in the presence of a
signal requires a prediction of its dilution by the background.  The background
is also needed in order to predict the size of the fluctuations of the SM
asymmetry around zero.

Direct use of Monte Carlo simulation to model the SM background
distribution would be a source of large systematic errors, as NLO
corrections are not known, and since we impose a hard cut on $S_T$.
We therefore propose a (partially) data-driven method, minimizing
this systematic error while keeping the statistical errors under
control.  The result can then be combined with a signal Monte Carlo to
predict the differential asymmetry in $M_{j1bW}$. The search for a
signal will then involve fitting this expectation to the data.

Our first task is to discuss how to obtain the prediction (which we will refer
to as a ``template'') for the
differential asymmetry in $M_{j1bW}$, under both the SM and NP
assumptions.  We will begin by arguing that the SM asymmetry template is
zero to a sufficiently good approximation.  Next we will make a
proposal for a partially data-driven method to determine the template for a given NP
assumption, with low systematic uncertainty.
Finally, we will estimate the sensitivity of our variables, using a
simplified statistical analysis based in part on our proposed method.  Along the
way we will find the preferred value of the $S_T$ cut.

\subsection{The SM Template: Essentially Zero\label{SMas}}

It is crucial for our measurement that the asymmetry in the SM background
be known, so that the presence of a signal can be detected.  It would be even better if the SM
asymmetry is very small.  Here we give evidence that this is indeed the case.

It is essential to recognize that the SM background to the $t\bar tj$ process is
very different from the SM background to the $t\bar t$ process.  In $t\bar t$, all
asymmetries are zero at LO.  The non-vanishing SM asymmetry in $t\bar{t}$ therefore arises from an NLO
effect, involving both virtual corrections to $t\bar t$ and real
emission, that is, $t\bar tj$.  The asymmetry therefore cannot be
studied at all with a leading-order event generator, and in a matched
sample (which contains $t\bar tj$ but not the virtual correction to $t\bar t$) it
would actually have the wrong sign.

However, for $t\bar tj$ itself, differential charge asymmetries at
LO are {\it not} zero.  The correction to these asymmetries from NLO
corrections to $t\bar tj$ are subleading in general. Therefore
we can ask the following question of an LO generator: although the generic observable
in $t\bar t j$ events will show a charge asymmetry, is this the case
for the $M_{j1bW}$ variable, or is any asymmetry washed out?

We find that the asymmetry in the mass variable is consistent with
zero, as one can see in Fig.~\ref{MassBGas}.  This also turns out to be true for the angle variable
which we will discuss later.  We emphasize that this was not
guaranteed to be the case.  One can find variables that, at LO and
at parton-level, exhibit asymmetries.  An example is the asymmetry
between the $P_T$ of the $t$ and that of the $\bar t$, which is of
order 4\% at parton-level. The fact that $qg\to t\bar tj$ has rather
small asymmetries, and that the symmetric $gg$ initial state
contributes significantly to $t\bar tj$, helps to reduce the size of
any observable asymmetries.  After reconstruction and detector
effects, nothing measurable remains.

\begin{figure}[t!]
\centering
\subfigure[$M_{j1bW}$ for SM background with a 700 GeV $S_T$ cut.]{
   \includegraphics[width =0.45\textwidth] {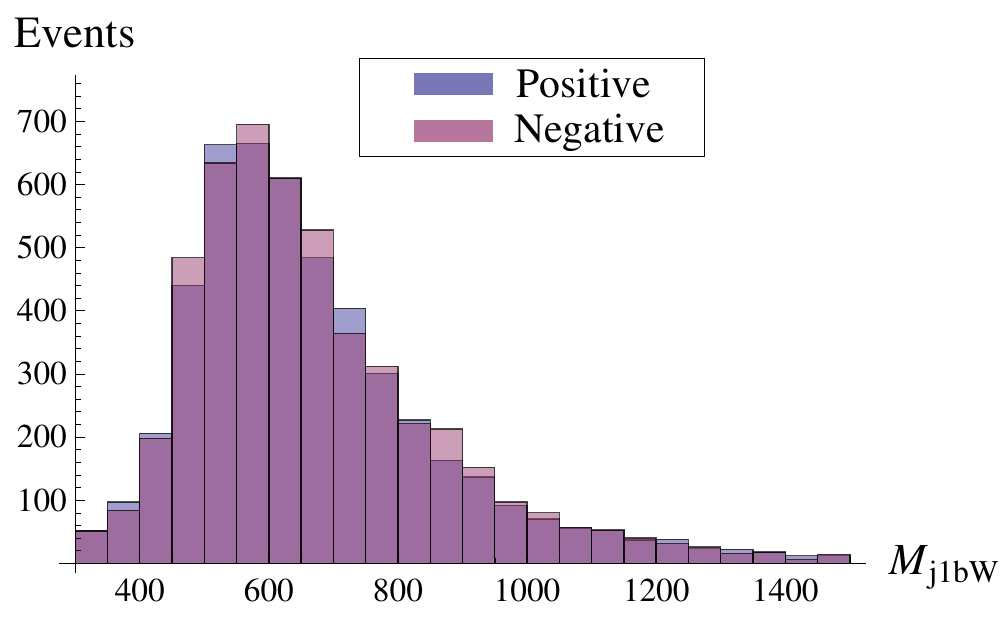}
   \label{bgmasslhe}
 }\hfill
 \subfigure[Bin-by-bin asymmetry in $M_{j1bW}$ for SM background with a 700 GeV $S_T$ cut.]{
   \includegraphics[width =0.48\textwidth] {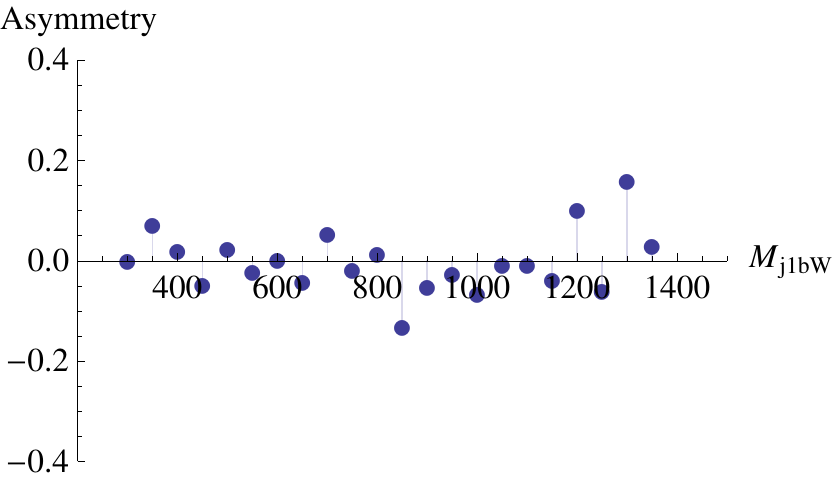}
   \label{bgmassbinlhe}
 }
\caption{A parton-level study of the SM background asymmetry for the mass variable with a 700 GeV $S_T$ cut,
corrsponding to 12 $\mathrm{fb}^{-1}$ luminosity. Other simulations confirm that the asymmetry appearing at 1200 -- 1400 GeV is a statistical fluctuation. \label{MassBGas}}
\end{figure}

We know of no reason why NLO
corrections would change this conclusion.  Neither virtual corrections nor real
jet emission have any reason to strongly affect $M_{j1bW}.$  For this
reason we will treat the SM background as purely symmetric.

No argument of this type is airtight.  Fortunately, the experiments do not need to
rely entirely upon it.  As we see in Fig.~\ref{wplhco}, the
asymmetry in the signal has a characteristic kinematic structure.  Moreover,
related asymmetries
will show up in several mass variables in a correlated way, due to the $W'$, and one would
not expect similar correlations in the background.  Finally, a signal is likely also
to appear in the angle variable discussed in Sec.~\ref{Secangle}.  The existence of these multiple cross-checks should allay any concerns that a measurement of a non-zero asymmetry might be uninterpretable.

\subsection{Obtaining NP Templates and Accounting for Fluctuations\label{secTemplate}}

We now discuss how to obtain the NP template
that is needed for each benchmark point.  In addition one needs to be able
to estimate the fluctuations that can occur under both the SM and NP
assumptions.  We emphasize the possibility of data-driven approaches.

\begin{table}[b]
\begin{tabular}{|c|p{13cm}|}
\hline
$B^+_i\:(B^-_i)$&Number of positive (negative) lepton events in $i^{th}$ bin, for background-only Monte Carlo.\\\hline
$S^+_i\:(S^-_i)$&As above, for signal-only Monte Carlo.\\\hline
$D^+_i\:(D^-_i)$&As above, in observed data.\\\hline
$[D^+_i]\:([D^-_i)]$&As above, in a fit to the observed data.\\\hline
$\hat A_i$&Predicted charge asymmetry the $i^{th}$ bin for a particular hypothesis.\\\hline
$ A_i$&Charge asymmetry in  $i^{th}$ bin as observed in data.\\\hline
$ c_n$&Amplitude for best fit of an NP template to the $n^{th}$ pseudo-experiment under the SM hypothesis.\\\hline
$ \tilde c$&Amplitude for best fit of an NP template to the asymmetry observed in the data.\\\hline
$ \sigma_c$&Standard deviation of the $c_n$.\\\hline
\end{tabular}
\caption{Notation used throughout Sec.~\ref{Secstat}.\label{notation}}
\end{table}

We will find it useful to introduce some notation (summarized in Table
\ref{notation}) in which $S^\pm_i$ and $B^\pm_i$ represent, for a
signal-only and background-only Monte Carlo sample, the number of
events in bin $i$ with a positively- or negatively-charged lepton
$\ell^\pm$.   $D^\pm_i$ denotes the similar quantity in data
(and is thus not generally equal to the expected result
$S^\pm_i+B^\pm_i$.)  At some point we will need a smoothed version of the data, which we denote via $[D^\pm_i]$.  The differential charge asymmetry predicted by
the template for a particular benchmark point, or by the SM itself, we
denote by $\hat A_i$.  Meanwhile, we call the observed asymmetry in
the data $A_i$.

Let us first focus on the statistical fluctuations around the template for the SM, which as 
we argued above in Sec.~\ref{SMas} can be taken to be zero.  Whenever one needs this template, it is under the assumption that the data is pure SM.  Even without signal, there will be plenty of data with $\geq$ 5 fb$^{-1}$ and an $S_T$ cut of order 700 GeV.  It therefore appears that rather than obtain the fluctuations around zero using a Monte Carlo sample $B_i$, one would have much smaller systematic errors using the data $D_i=D_i^++D_i^-$ itself.  One could probably do even better using a fit $[D_i]$ to the data, smoothing the bin-by-bin fluctuations in the numbers of events. We believe that the remaining statistical uncertainties that come with this method of modeling background will be smaller than the systematic uncertainties on an LO Monte Carlo for $B_i$. From this data-driven
 model, one may determine the expected size of the fluctuations on $\hat A_i$ by performing a series of pseudo-experiments.

Next let us consider how to determine the template $\hat A^i$ for a particular NP hypothesis 
We could of course simply compute it from large Monte Carlo samples, with Monte Carlo integrated luminosity $\mathcal{L}_{MC}$ much larger than the integrated luminosity in data $\mathcal{L}_{data}$, for $S_i$ and $B_i$.
\begin{equation}\label{MCtemplate}
\hat A_i \equiv\frac{S_i^+-S_i^-}{S_i^++S_i^-+B_i^++B^-_i}
\end{equation}
(Recall we are ignoring interference for now.\footnote{If
interference cannot be neglected, as might happen with very large data sets or perhaps
with other models that we have not explored in detail,
then our separation of $S_i$ and $B_i$ is naive. What must then appear in the numerator is the
difference of positive and negative lepton events in the combined signal and background.  Systematic
errors will then presumably be somewhat larger.})  Here the $B_i^\pm$ cancel in the numerator, since the asymmetry in the SM background
is assumed to be zero. With this approach statistical errors can be made
arbitrarily small, but systematic errors on the SM background prediction could
be very substantial.  The process $t\bar tj$ has never previously been measured at 
these energies, and after the $S_T$ cut it is difficult to estimate how large the systematic 
errors might be.  Moreover we know of no way to extract the $t\bar tj$ background reliably, in the presence of signal, without the potential for signal contamination. 

An alternative purely data-driven approach would be to use the suitably-fitted charge-symmetric data $[D_i^++D_i^-]$ in the denominator of (\ref{MCtemplate}). For the numerator one may take a large Monte Carlo sample for $S_i$, and scale it to the luminosity of the data sample, giving
\begin{equation}\label{datatemplate}
\hat A_i
\equiv\frac{(S_i^+-S_i^-)\frac{\mathcal{L}_{data}}{\mathcal{L}_{MC}}}{[D^+_i+D^-_
i]}
\end{equation}
where again $\mathcal{L}_{data}$ and $\mathcal{L}_{MC}$ are the luminosities of the
data and the signal Monte Carlo sample.  This method introduces correlations between the prediction of the template $\hat A^i$ and the measurement $A^i$ which would have to be studied and accounted for.  However, the systematic error introduced by these correlations may in many cases be much smaller than those introduced by relying on a Monte Carlo simulation for the denominator, as in (\ref{MCtemplate}).  In addition, statistical errors that arise from the finite amount of data, which would be absent with a large Monte Carlo sample, are negligible, as can be seen as follows.   The statistical error on the {\it predicted asymmetry} $\hat A_i$ is dominated by fluctuations of the denominator of (\ref{datatemplate}), since the statistical error on the {\it numerator} of (\ref{datatemplate}) can be made arbitrarily small by increasing $\mathcal{L}_{MC}$:
\begin{equation}
\frac{\sigma(\hat A_i)}{\hat A_i}= \frac{1}{[D^+_i+D^-_
i]^{1/2}}.
\end{equation}
However, for the {\it measured asymmetry} $A_i$, defined as $A_i\equiv\frac{D^+_i-D^-_i}{D^+_i+D^-_i}$, the
error is always (for these models) dominated by the numerator:
\begin{equation}
\frac{\sigma(A_i)}{A_i}= \frac{1}{(D^+_i+D^-_
i)^{1/2}}\sqrt{1+\frac{1}{A^2_i}}.
\end{equation}
More precisely, since the largest observed asymmetries per bin will be of the order of $0.15$, the
statistical error on the observed asymmetry is always larger than the
statistical error on the template --- $\sigma(A_i)>>\sigma(\hat A_i)$. And again we emphasize that this data-driven method
reduces systematic uncertainties from what is often the largest source: the lack of confidence that the $t\bar tj$ background is correctly modeled.  This comes at the relatively low cost of mild correlations between prediction and data, and some additional minor statistical uncertainty.

Partially data-driven approaches are also possible.  Even if one uses $[D_i]$, the choice of fitting function could be determined in part with the use of Monte Carlos for $B_i$ and $S_i$.  Interestingly, the distribution in the variable $M_{j1bW}$ is quite similar in signal and background, so the presence of signal, though it affects the overall rate, does not strongly affect the overall shape away from the $W'$ resonance.  

Since the pros and cons of these methods are luminosity-dependent, and dependent upon the details of the analysis, the only way to choose among these options is to do a study at the time that the measurement is to be made.  We therefore do not attempt any optimization here. Whatever method is used, the last step in the process in obtaining the NP template is to fit the $\hat A_i$ to a smooth function, which then serves
as the template for the asymmetry in this particular benchmark point.  (The size of the fluctuations around this template can again be obtained from $[D_i]$, as we suggested for the SM template.)
After repeating this process for a grid of benchmark points, one may then compare the data to the SM null template or to any one of the NP templates.  In the next subsection we will carry out a simplified version of this study, to investigate the effectiveness of our methods.

\subsection{Effectiveness of Our Method: A Rough Test\label{resultsSec}}

A full evaluation of our method,
carrying out precisely the same analysis that the
experimentalists will need to pursue, would require more firepower
than we have available. Instead we will
carry out a somewhat simplified analysis, asking the
following question:

\begin{center}\emph{If the NP hypothesis for a certain benchmark point is realized in the data, what is the average confidence level at which we can reject the SM hypothesis?}\end{center}

\noindent The answer to this question will serve two purposes.  First, it will give
a measure of how sensitive a complete analysis will be for distinguishing the SM from various NP scenarios.  (More
precisely, it will be slightly optimistic, as we will discuss, but not overly so.)
Second, it will allow us to estimate what value of the $S_T$ cut is
optimal for different benchmark points.

We have not yet said much about the $S_T$ cut, so let us remark on it now.
Without such a cut, the signal to background ratio in the $t\bar tj$ sample is small, as small
as 1:45 for $m_{W'}=800$ GeV with $g_R={\sqrt{2}}$.  However, the
situation can be much improved using the fact that the signal $S_T$
distribution tends (especially for heavy $W'$s) to sit at much larger values for
signal than for the SM background. (See Fig.~\ref{stdist}; note these plots show the $S_T$ distributions for our large-$g_R$ benchmark points.  From this one can see that a simple counting experiment would not be trivial.)  The optimal value of the $S_T$ cut
depends on the model, the analysis method and the luminosity.  For most of our
purposes an $ S_T$ cut of the order of 700 GeV is suitable, as we will see later.

\begin{figure}[t!]
\centering \subfigure[$M_{W'}=400\:\mathrm{GeV},\:g_R=1.5$]{
   \includegraphics[width =0.3\textwidth] {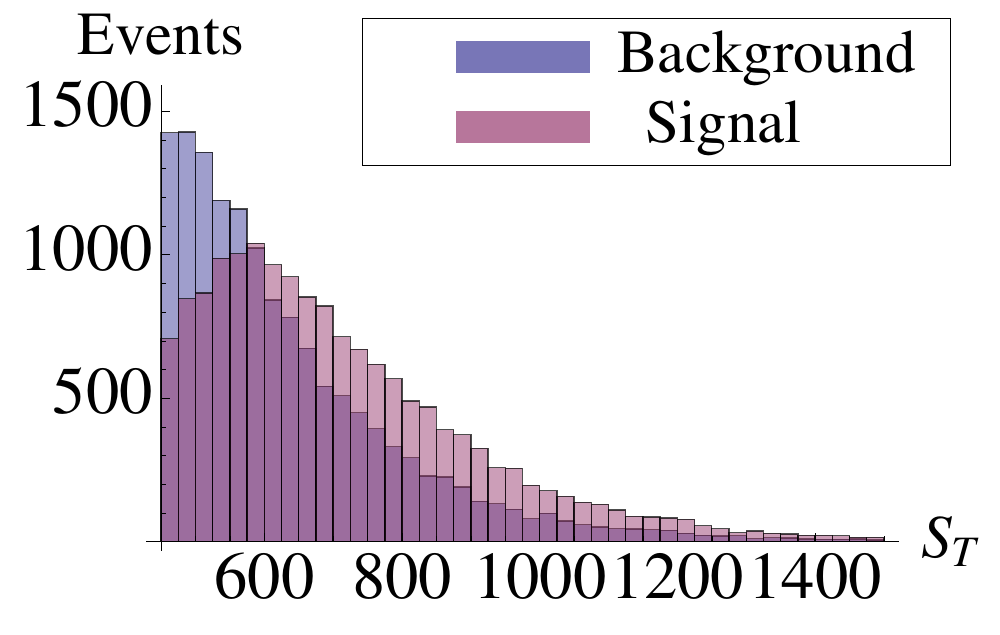}
   \label{st1}
 }\hfill
 \subfigure[$M_{W'}=600\:\mathrm{GeV},\:g_R=2$]{
   \includegraphics[width =0.3\textwidth] {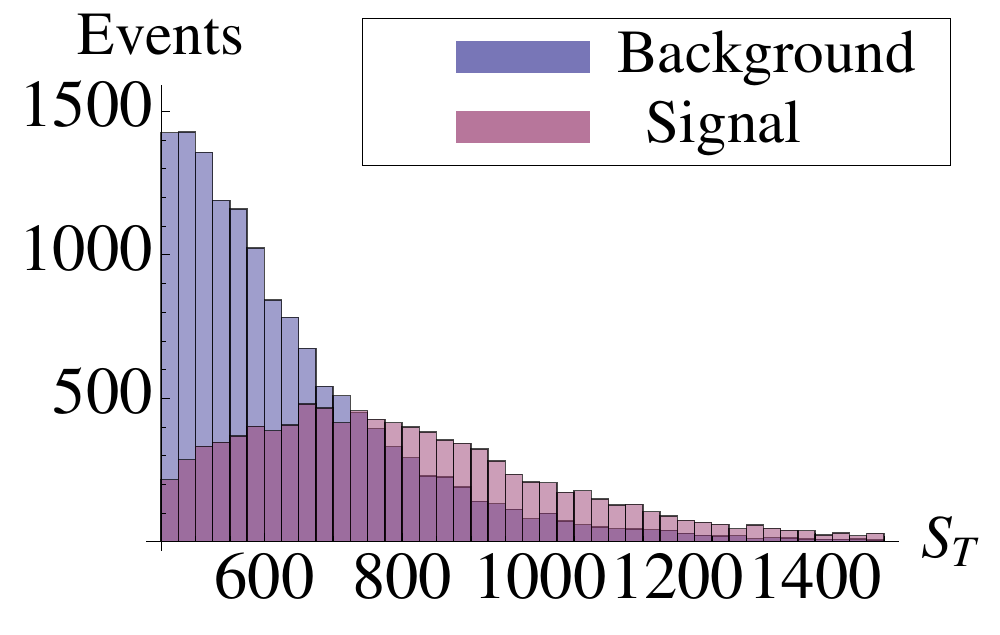}
   \label{st2}
 }\hfill
 \subfigure[$M_{W'}=800\:\mathrm{GeV},\:g_R=2$]{
   \includegraphics[width =0.3\textwidth] {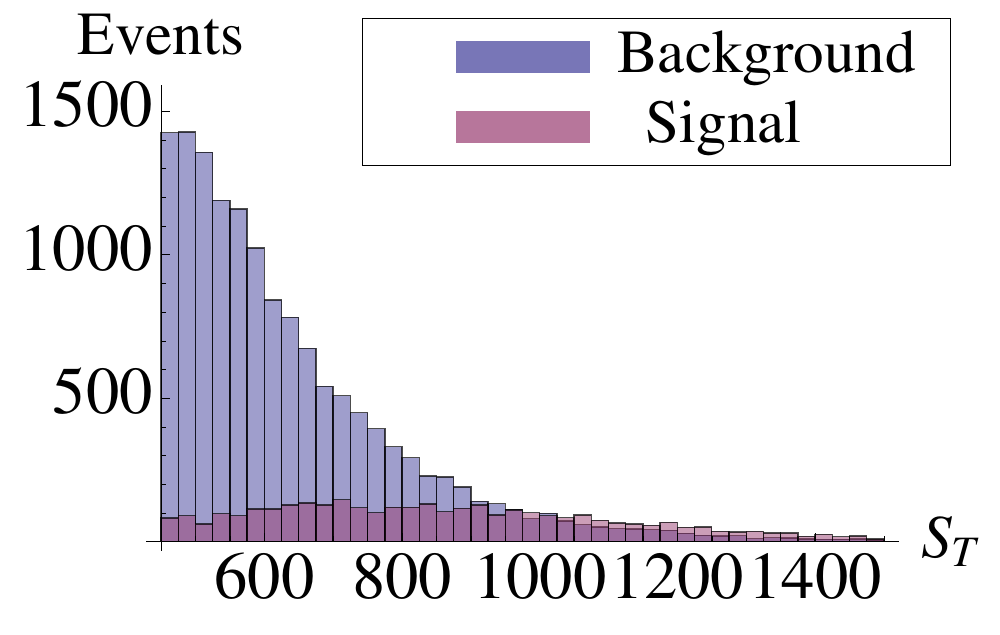}
   \label{st3}
 }

\caption{$S_T$ distributions of signal and background, for various
benchmark points. For computational reasons we did not simulate
events with $S_T<450$ GeV. The samples correspond to 8 fb$^{-1}$.
\label{stdist}}
\end{figure}

Answering the italicized question posed above is equivalent to
evaluating the probability for fluctuations about the SM assumption to create a differential asymmetry $A_i$
that resembles the pattern predicted by the NP
assumption $\hat A_i$. For this we need (a) the template $\hat A_i$ for the NP assumption and
(b) an estimate of the size of the fluctuations that can occur under
the SM assumption.

We have discussed above how to obtain these things from the data at the LHC.
But since the actual data $D_i$ are not yet available, we obtain our
NP template $\hat A_i$ from large $S_i$ and $B_i$ Monte Carlo samples, using formula (\ref{MCtemplate}).
Obtaining the fluctuations under the SM assumption is a bit subtle.
Since in this section we are {\it assuming} the data itself contains a signal, our background model must be obtained,
according to our data-driven strategy, from our simulation of $S_i+B_i$ (and not from $B_i$ alone!)  We take
the expected numbers of positive- and negative-charge lepton events
to both be equal to half of $S_i+B_i$.  We then study
the fluctuations around this background model by performing 50 000 Poisson-fluctuating
pseudo-experiments, for positive- and negative-lepton events independently,
and computing the differential asymmetry for each one.

Finally, to address our italicized question, we must then ask: what
is the probability for fluctuations of the asymmetry  around zero,
given this background model, to resemble the ``data''?  This is done
as follows: For each pseudo-experiment, we fit the differential
asymmetry to the NP template $\hat A_i$ of our benchmark point,
\emph{keeping the shape of the NP template fixed but allowing the
amplitude to float}. The best-fit amplitude we denote by $c_n$,
where the index $n$ labels the pseudo-experiment.  For illustration,
some examples for a couple of pseudo-experiments are shown in
Figs.~\ref{pseudo1} and \ref{pseudo2}.

\begin{figure}[t!]
\centering \subfigure[SM pseudo-experiment, with amplitude $0.39$]{
   \includegraphics[width =0.3\textwidth] {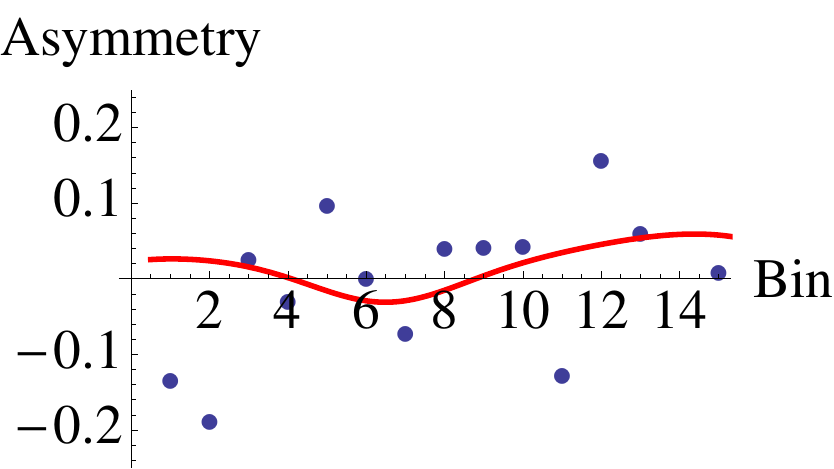}
   \label{pseudo1}
 }\hfill
 \subfigure[SM pseudo-experiment,  with amplitude $-0.16$]{
   \includegraphics[width =0.3\textwidth] {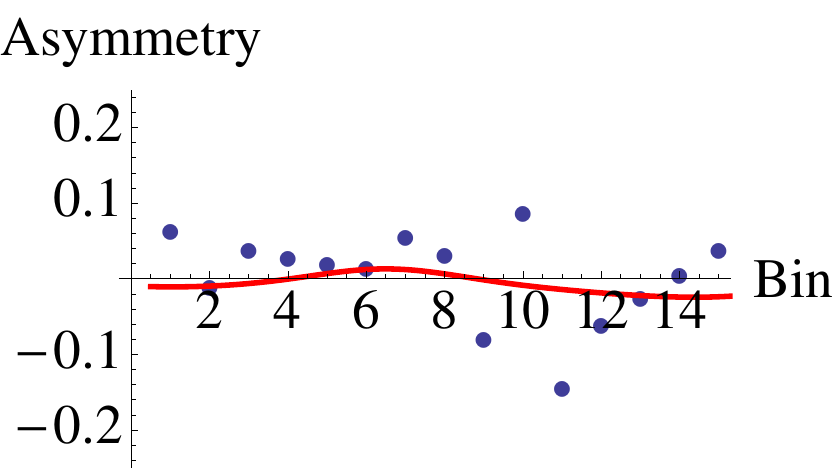}
   \label{pseudo2}
 }\hfill
 \subfigure[NP pseudo-experiment,  with amplitude $0.98$]{
   \includegraphics[width =0.3\textwidth] {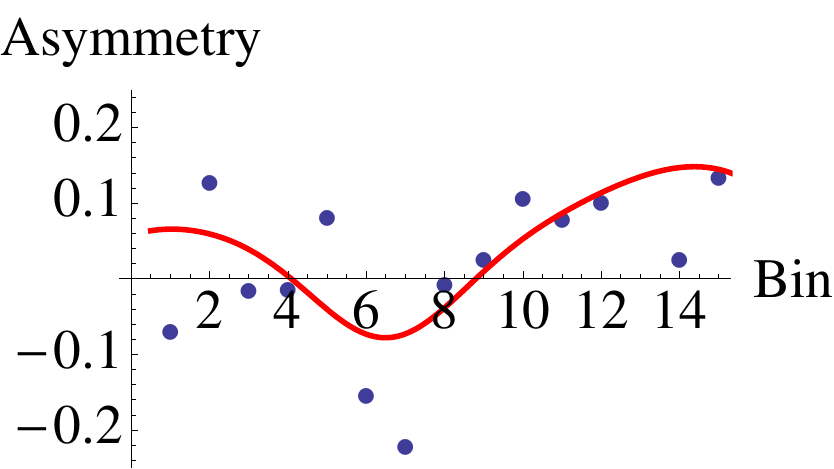}
   \label{pseudo3}
 }

\caption{Two examples of possible fluctuations of the differential
charge asymmetry under the SM hypothesis, and one example under the
NP hypothesis. The red line is the best fit of the amplitude of the
NP template to the pseudo-experiment, with the shape held fixed. The
NP template that was chosen corresponds to the 600 GeV $W'$ with
$g_R=2$ and $S_T>700$ GeV. The fluctuations are representative for a 5
$\mathrm{fb}^{-1}$ sample. \label{pseudoexp}}
\end{figure}

Under the SM assumption (zero asymmetry), the expectation value  of the $c_n$ is zero.  (Similarly, under the correct NP assumption, the expectation would be 1.)
The $c_n$ follow a Gaussian distribution, whose width gives the standard deviation $\sigma_c$ of the $c_n$ around zero. If an amplitude of size $\tilde c$ were observed in the data, the $p$-value (chance of a fluctuation on the SM hypothesis to produce a structure with amplitude $\tilde c$ or larger) is then:
\begin{equation}
P[X>\tilde c]=\frac{1}{\sqrt{2\pi}\sigma_c}\int^{+\infty}_{\tilde c}\!\!\!\!\!\!dc\:e^{-\frac{1}{2}(c/\sigma_c)^2}.
\end{equation}
To get a measure of typical significance, we compute $P[X>1]$, the
probability for the SM to produce an $A_i$ resembling the template
$\hat A_i$ with an amplitude $\tilde c$ exceeding 1. (Recall that
$\tilde c=1$ would be the expected value given that nature has chosen
this benchmark point.)  The results of this procedure for our
benchmark points, after conversion to standard deviations on a
Gaussian, are displayed in Table \ref{tableresults}, for two
integrated luminosities and for the optimal $S_T$-cut (see below.)
In Appendix~\ref{appplots}, we also present contour plots of the
significance as a function of the integrated luminosity and the
$S_T$ cut; see Figs.~\ref{contourplots1} and \ref{contourplots2}.

The amount by which the
{\it observed} significance tends to fluctuate around the {\it expected}
significance depends on the luminosity and the
$S_T$ cut. By running a different set of pseudo-experiments based on the NP hypothesis, we can obtain the Gaussian
distribution of the amplitude of the fit.  (An example of such a pseudo-experiment is shown in Fig.~\ref{pseudo3}.)
Values for the width of this distribution give us the statistical error bar on the expected significance, and are included in Table \ref{tableresults}.

\begin{table}[t]
\centering
\renewcommand{\arraystretch}{1.2}
\begin{tabular}{|c|c|c||c|c|}
\hline
\multirow{2}{*}{\begin{tabular}{c}$ \quad M_{W'}\quad$\\[-1.5ex] (GeV)\end{tabular} }&\multirow{2}{*}{$\quad g_R\quad$}&\multirow{2}{*}{\begin{tabular}{c}$\quad S_T$ cut$\quad$\\[-1.5ex] (GeV)\end{tabular} }&\multicolumn{2}{c|}{ $\quad$Significance $\quad$}\\\cline{4-5}
&&&$\quad \quad 5\: \mathrm{fb}^{-1}\quad \quad $&$\quad \quad 8\: \mathrm{fb}^{-1}
\quad\quad$\\\hline\hline
400  & 1.5 &750&$6.27\pm0.92$&$7.49\pm0.75$\\
400  & $\frac{1.5}{\sqrt{2}}$ &750&$3.38\pm0.95$&$4.24\pm0.95$\\
600  & 2&700&$3.42\pm0.92$&$4.08\pm0.95$\\
600 & $\sqrt{2}$&700&$1.79\pm0.83$&$2.15\pm0.86$\\
800 & 2 &700&$2.37\pm0.87$&$3.12\pm0.92$\\
800 & $\sqrt{2}$ &700&$1.60\pm0.82$&$2.01\pm0.85$\\[0.7ex]
\hline
\end{tabular}
\caption{Expected significance and statistical error for SM exclusion at our benchmark points, given selected luminosities and optimal $S_T$ cuts. For the correct interpretation of these numbers, please refer to the text. \label{tableresults}}
\end{table}

Our simplified analysis is imperfect in various ways.  One important weakness is that we assume that nature matches one of our benchmark points, and
we do not consider the effect of using the wrong benchmark point in obtaining the exclusion of the SM.
In particular, the mass of the $W'$ we used to obtain the NP template matches the mass of the $W'$ in our ``data''.   A finer grid in $W'$ mass would address
this. 
(In general, the coupling $g_R$ for the template will also differ from
the real coupling, but except for its effect on the $W'$
width, often smaller than the experimental resolution, a change in
the coupling affects the amplitude, but not the shape, of the
corresponding template.) Also, our simplified procedure to fit only for the amplitude
of the template and to keep the shape fixed does not always capture
all the features of the asymmetry distribution, as is illustrated in
Fig.~\ref{pseudo3}, where the central dip in the asymmetry is deeper
than our fit function can capture. In a more detailed study one
might choose to let multiple parameters float to obtain a better
fit. We further note that we are not accounting for the look-elsewhere effect. 
And finally, although the use of asymmetries and a data-driven method reduces systematic errors, we have not considered the remaining systematic errors here.

On the other hand, there are important features of the signal that we are not using in our analysis, and including those would enhance the sensitivity.  The use of several (correlated) mass variables, and the angle variable discussed in the next section, would give some improvements.  Moreover, while the charge asymmetry we focus on here has low systematic errors but is statistically limited, other observables with higher systematics but lower statistical errors, such as the differential cross-section with respect to $S_T$,  are obviously useful as well. In any search for this type of models multiple approaches should be combined. 

\section{An angle variable\label{Secangle}}
In this section we discuss another charge-asymmetric variable,
the azimuthal angle between the
hardest jet without a $b$-tag ($j_1$) and the lepton $\ell$:
\begin{equation}
\Delta \phi_{j_1,\ell}=\mathrm{Min}\Big[|\phi_{j_1}-\phi_{\ell}|,\:2\pi-|\phi_{j_1}-\phi_{\ell}|\Big].
\end{equation}

With a low $S_T$ cut, the angle between the hardest jet and an
$\ell^-$ tends to be larger than the angle between the hardest jet
and an $\ell^+$ [Figs.~\ref{PhiDiffSLow} and \ref{PhiAsymmlow}]. The
reason is as follows: The $W'$ is produced near threshold, so the
recoiling top quark or antiquark is not highly boosted. The top from the $W'$
decay, on the other hand, will recoil back-to-back against the $d$
or $\bar d$ (which is usually the source of the hardest jet).
Moreover, this top will be somewhat boosted since $m_{W'}\gg m_t$,
so if it decays leptonically, the lepton's momentum tends also to be
back-to-back to the $d$ or $\bar d$. This results in a large opening
angle between the hardest jet and the lepton. However, if it is the
other top quark that decays leptonically, the angle of its lepton
with the hardest jet is more randomly distributed. Since the
negatively charged $W'$ is produced more abundantly, this variable
will exhibit a charge asymmetry.

For a high $ S_T$ cut the picture reverses. The $W'$ and the top
from which it recoils are now both boosted and typically
back-to-back with one other. The decay products from the $W'$ tend
to be aligned with each other.  In other words, a cluster of four
objects (from the $W'$) is now recoiling against a cluster of three
objects (the top). The hardest jet is typically still the down quark
from the $W'$ decay.  If the lepton's parent is the top from the
$W'$, $\Delta\phi_{j_1\ell}$ tends to be small, while the opposite
is true if the lepton comes from the recoiling top. (See
Figs.~\ref{PhiDiffSHigh} and \ref{PhiAsymmhigh}.)

\begin{figure}[t]
\centering \subfigure[The angle variable in signal only for $M_{W'}=800$ GeV and $g_R=2$.]{
   \includegraphics[width =0.45\textwidth] {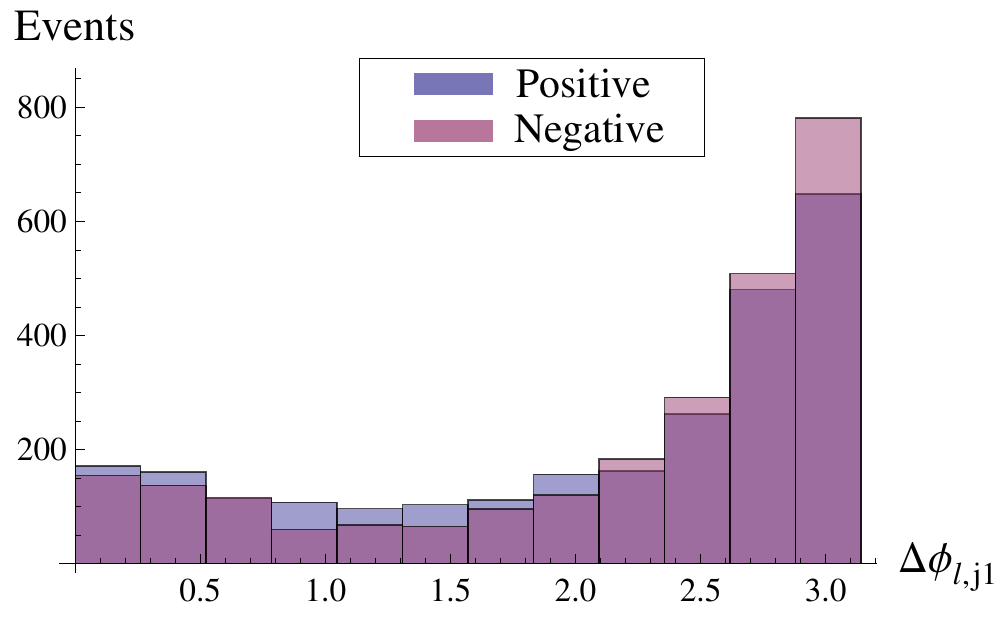}
   \label{PhiDiffSLow}
 }\hfill
 \subfigure[Asymmetry of the angle variable in signal only for $M_{W'}=800$ GeV and $g_R=2$.]{
   \includegraphics[width =0.48\textwidth] {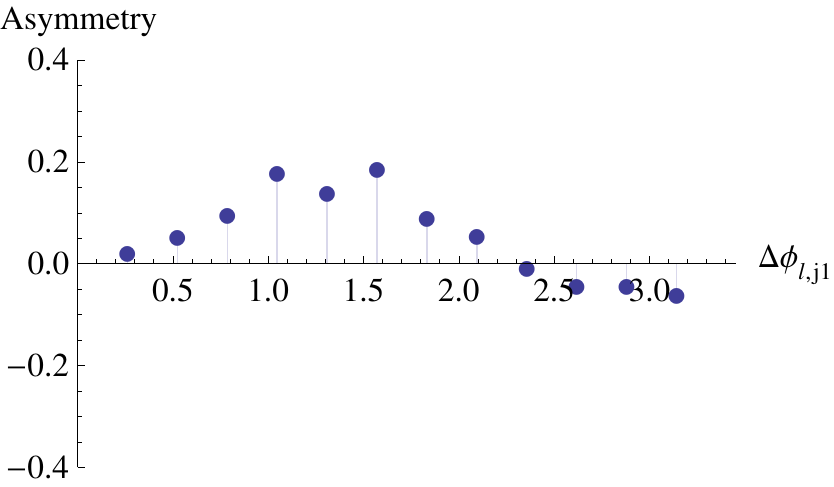}
   \label{PhiAsymmlow}
    }\hfill
 \subfigure[The angle variable in signal only for $M_{W'}=400$ GeV and $g_R=1.5$.]{
   \includegraphics[width =0.45\textwidth] {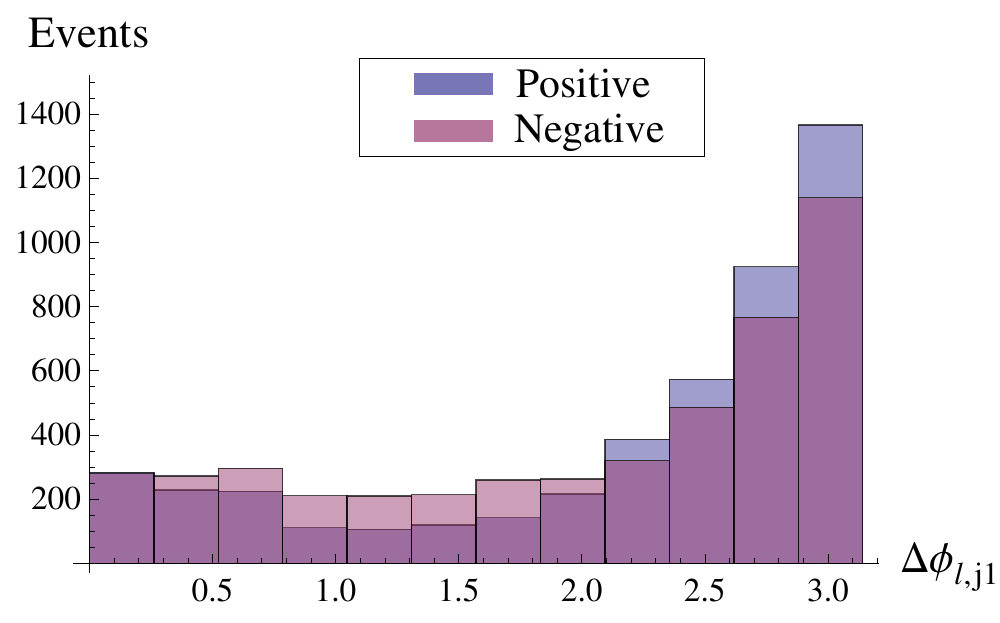}
   \label{PhiDiffSHigh}
    }\hfill
 \subfigure[Asymmetry of the angle variable in
signal only for $M_{W'}=400$ GeV and $g_R=1.5$.]{
   \includegraphics[width =0.48\textwidth] {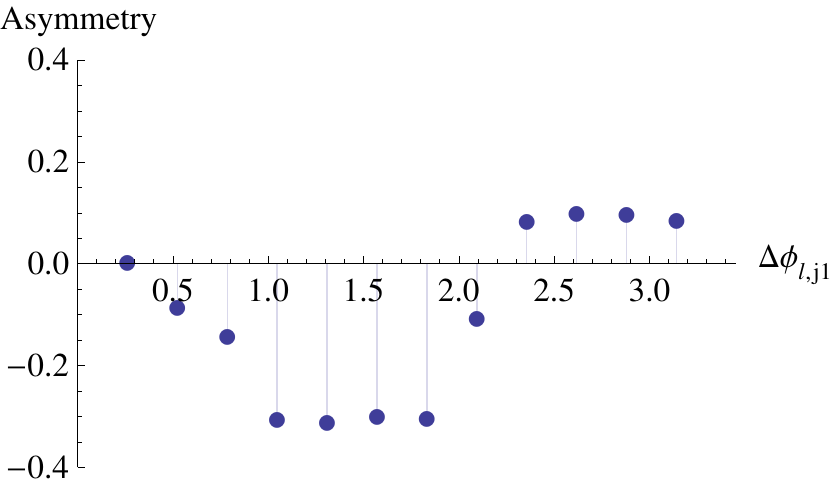}
   \label{PhiAsymmhigh}
  }

\caption{Angle difference between the lepton and the hardest jet at parton-level signal-only, for $W'$s of mass 800 and 400 GeV with an
$S_T$ cut at 700 GeV. The samples correspond to 5 fb$^{-1}$. }
\end{figure}

This
reversing structure in the asymmetry as a function of the $S_T$ cut is useful,
as it potentially provides a very strong hint of
new physics.  However, there is an intermediate $S_T$ cut where the asymmetry is essentially
zero, so in that range the variable is not useful.  For this reason, we recommend studying this
variable {\it as a function of the $S_T$ cut}.

We explicitly checked that the standard model will not introduce a large asymmetry in this angle variable, for any $S_T$ cut.  A particular case is shown in Fig.~\ref{BGASYMM}.  Our reasoning for
trusting a LO Monte Carlo is the same as was described in Sec.~\ref{SMas} for the mass variable.

\begin{figure}[h]
\centering
 \subfigure[The angle variable for SM background with a 700 GeV $S_T$ cut.]{
   \includegraphics[width =0.43\textwidth] {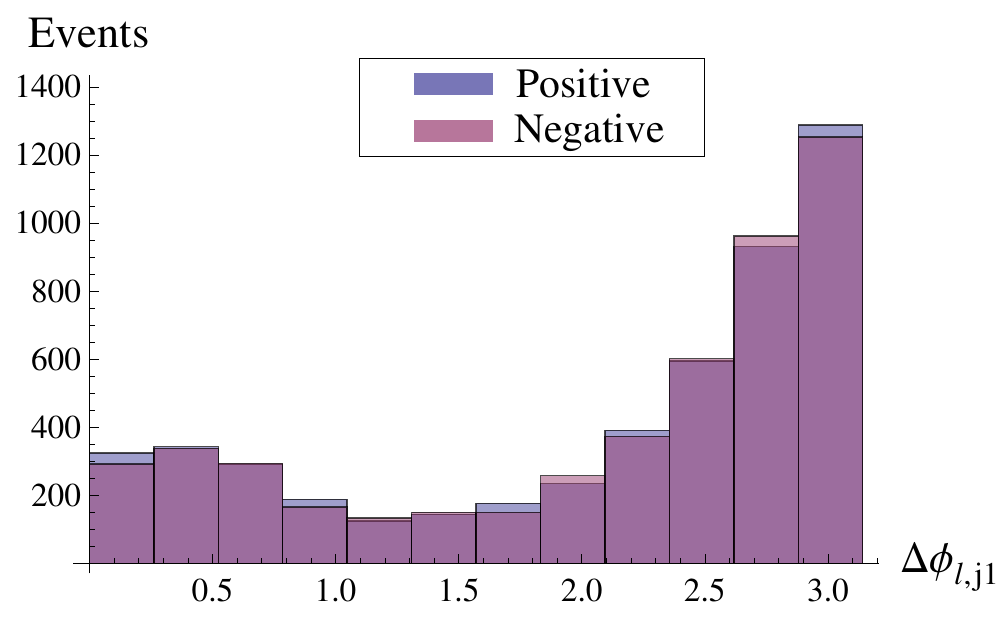}
   \label{bganglhe}
 }\hfill
  \subfigure[Bin-by-bin asymmetry in the angle variable for SM background with a 700 GeV $S_T$ cut.]{
   \includegraphics[width =0.48\textwidth] {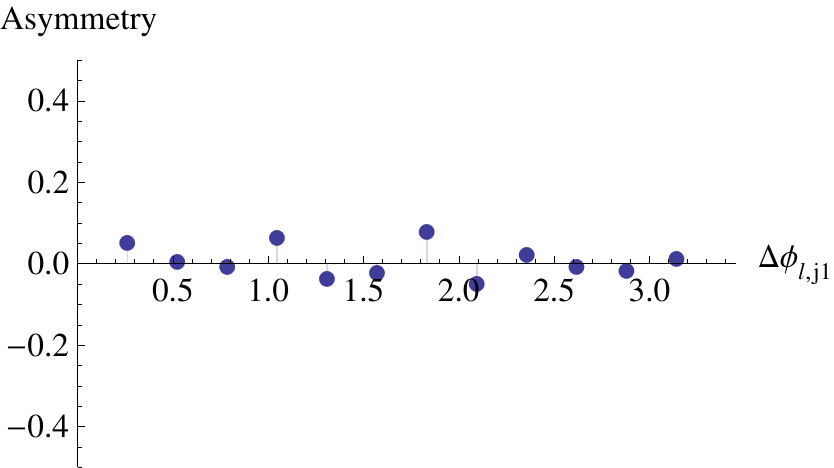}
   \label{bgangbinlhe}}
\caption{A parton-level study on SM background asymmetry for the angle
variable with a 700 GeV $S_T$ cut, corresponding to 12 fb$^{-1}$
luminosity.\label{BGASYMM}}
\end{figure}

An interesting feature of this angle variable in the $W'$ model
(though whether this is true in other models has not yet been
studied) is that the point where the number of positive and negative
lepton events is roughly equal is insensitive to $m_{W'}$ and $g_R$.
For all our benchmark points we find $\Delta\phi_{j_1,\ell}\approx
2$ to be a suitable place to break the signal into two bins.  The
detector-level asymmetries in both bins are given in Table \ref{angletableresults}.
To estimate the significance, we follow a strategy similar to the
one mentioned for the mass variable. However instead of fitting for
the amplitude of a previously obtained template, we compute the
difference of the asymmetry of the two superbins and establish the
Gaussian probability distribution for this variable using
pseudo-experiments on the SM hypothesis.  Plots of the resulting
significance of this observable as a function of $S_T$ cut and
luminosity can be found in Figs.~\ref{contourplots3} and
\ref{contourplots4} in Appendix~\ref{appplots}.

\begin{table}[t]
\centering
\renewcommand{\arraystretch}{1.2}
\begin{tabular}{|c|c|c||c|c|}
\hline
\multirow{2}{*}{\begin{tabular}{c}$ \quad M_{W'}\quad$\\[-1.5ex] (GeV)\end{tabular} }&\multirow{2}{*}{$\quad g_R\quad$}&\multirow{2}{*}{\begin{tabular}{c}$\quad S_T$ cut$\quad$\\[-1.5ex] (GeV)\end{tabular} }&\multicolumn{2}{c|}{Asymmetry (\%)}\\\cline{4-5}
&&&$\quad1^{\mathrm{st}}$ bin$\quad$&$\quad2^{\mathrm{th}}$ bin$\quad$\\\hline\hline
400  & 1.5 &800&-13.7&10.2\\
400  & $\frac{1.5}{\sqrt{2}}$ &800&-9.3&7.0\\
600  & 2&1200&-9.6&12\\
600 & $\sqrt{2}$&1200&-6.8&8.7\\
800 & 2 &700&3.8&-2.4\\
800 & $\sqrt{2}$ &700&2.4&-1.7\\[0.7ex]
\hline
\end{tabular}
\caption{Expected asymmetry at detector-level in the angle variable
for each superbin, for our benchmark points using the optimal $S_T$
cut. \label{angletableresults}}
\end{table}

The greatest merit of the angle variable is its simplicity. Both the
hardest jet and the lepton are well-measured, and in contrast to the
mass variables no (partial) event reconstruction is needed.
Unfortunately the angle variable is more sensitive than $M_{j1bW}$ to interference
effects between signal and background.
Whether the contribution from interference is positive or negative
depends on the mass of new particle, the $S_T$ cut and the model we
study.  The effect, however, appears to be only moderate.
We find that, for a $W'$ mass of 800 GeV and an $S_T$ cut of 700
GeV, the asymmetry for the two bins after interference is included
is reduced by about 15\%. A more detailed study including
interference is advisable to give a precise estimate of its effects,
especially for other models where interference might be more important.

\section{Final Remarks}

At the LHC, models that attempt to explain the Tevatron $t\bar t$
forward-backward asymmetry with the exchange of a particle $X$ in
the $t$- or $u$-channel generate a charge-asymmetric signal in $tX$
production.  This leads to observable charge asymmetries in certain
variables within $t\bar tj$ samples.  Among interesting observables
are mass variables involving various final state objects including
the hardest jet and/or the lepton (Secs.~\ref{Secvar} and
\ref{Secstat}), the azimuthal angle between the lepton and the
hardest jet (Sec.~\ref{Secangle}) and the $P_T$ difference between
the tops and $W$ bosons (Appendix~\ref{AppPtdiff}). Of these
variables, the invariant mass of the hardest jet, the
leptonic $W$ and a $b$-tagged jet appears to be the most powerful and the most
universal, since it tends to reconstruct the $W'$ mass resonance.
The charge asymmetry of this variable exhibits a negative asymmetry
in the region of the $W'$ mass, and a positive asymmetry elsewhere.
We have proposed a data-driven method to extract a statistical
significance from this asymmetry structure.

One could of course go further by fully reconstructing the events, and directly observe that $W'^-$ production is larger than $W'^+$ production.  However demanding full reconstruction would lead to a considerable loss of efficiency.  Since we cannot realistically estimate this efficiency loss, we cannot evaluate the pros and cons of this approach, but clearly the experiments should do so.

We have described this asymmetry measurement on its own, without
discussing the fact that simultaneously the experiments will be
measuring charge-symmetric variables, such as the cross-section for
$t\bar tj$ as a function of $S_T$.  Of course these variables are complementary, and we do not in any
way mean to suggest that one should do one instead of the other.
Charge-symmetric variables may often have lower statistical uncertainties,
but in most cases background-subtraction is necessary, so there will
be large systematic errors.  The combination of the two types of
measurements will help clarify the situation far better than either
one could in isolation.  Additional information will come from the differential
charge asymmetry in $t\bar t$ events at the LHC, which is a direct test of the
Tevatron measurement of the $t\bar t$ forward-backward asymmetry,
and is sensitive to any growth of the effect with energy.

A very important aspect of our approach is that the asymmetry is a
diagnostic for models.  An $s$-channel mediator will not generate a
peak for either lepton charge, and so even if an asymmetry in $t\bar
tj$ were generated, it would be largely washed out in the variable
$M_{j1bW}$.  Among models with $t$- or $u$-channel mediators $X$,
{\it some will produce a negative asymmetry at $M_{j1bW}=m_X$, while
others will produce a positive asymmetry.}
For example, models that replace the $W'$ by a color triplet or color sextet scalar $X$ \cite{Gresham:2011pa,Shu:2011au,Gresham:2011fx,Westhoff:2011tq} that couples to $u_R$ and $t_R$ (and has charge 4/3) will have the opposite sign, because the process $u g \rightarrow \bar t X^+$ will be larger than $\bar u g \rightarrow t X^-$.  The approach we use will still apply, but the asymmetry will be positive in the neighborhood of the $X$ mass peak, rather than negative as it is for the $W'$.
For this reason, {\it even if it
turns out that the asymmetry measurement is not needed for a
discovery of the $X$ particle, it will still be an essential ingredient in
determining its quantum numbers and couplings.}

What seems clear from our results is that the data already available (or
soon to be available)
at the 7 TeV LHC should be sufficient to allow
for an informative measurement of charge-asymmetric observables in
$t\bar tj$ to be carried out.  We look forward to seeing studies of
$t\bar tj$ from ATLAS and CMS, and we hope that measurements of
charge asymmetries will be among them.

\begin{acknowledgements}
We would like to thank Sanjay Arora, John Paul Chou, Yuri Gershtein, Eva
Halkiadakis, Ian-Woo Kim, Amit Lath, Michael Park, Claudia Seitz,
Sunil Somalwar and Scott Thomas for useful discussions. We thank
Jiabin Wang for insights in the statistical procedure and we thank
Olivier Mattelaer for advice on the use of DELPHES.  The work of
S.K. and Y.Z. was supported by NSF grant PHY-0904069 and DOE grant
DE-FG02-96ER40959 respectively.  M.J.S. was supported by NSF grant
PHY-0904069 and by DOE grant DE-FG02-96ER40959.

\end{acknowledgements}

\clearpage


\appendix

\section{Additional Results\label{appplots}}
\subsection{Contour plots for the mass variable}
As can be seen in Figs.~\ref{contourplots1} and
\ref{contourplots2}, we
find that the optimal $ S_T$-cut for the mass variable does not vary greatly with luminosity, or even with the $W'$ mass: it
lies around $700$ GeV for the $600$ GeV and $800$ GeV $W'$ and is slightly
higher for the $400$ GeV $W'$. At lower $S_T$ cuts, reduced signal-to-background ratio worsens the significance. The reason a large $S_T$ cut works well even for low $W'$ mass is that the
distribution for the charge-symmetric component of the signal (mainly
$t$-channel $W'$ exchange) peaks at low $S_T$ for a lighter $W'$.
Meanwhile, for an overly
high $ S_T$ cut the remaining signal is too small.
But we should mention that our binning
procedure makes our results too pessimistic here.

\newcommand{\contsize}{0.38}

\begin{figure}[b]
\subfigure[$M_{W'}=400$ GeV, $g_R=1.5$]{
   \includegraphics[width = \contsize\textwidth] {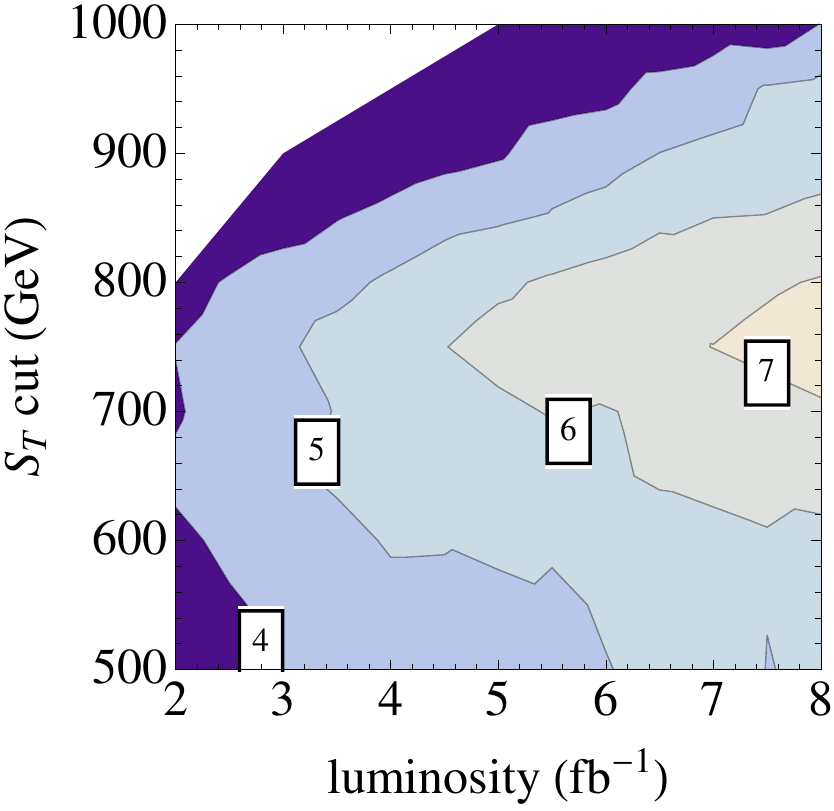}
   \label{wp400g}
 }\hfill
 \subfigure[$M_{W'}=400$ GeV, $g_R=\frac{1.5}{\sqrt{2}}$]{
   \includegraphics[width = \contsize\textwidth] {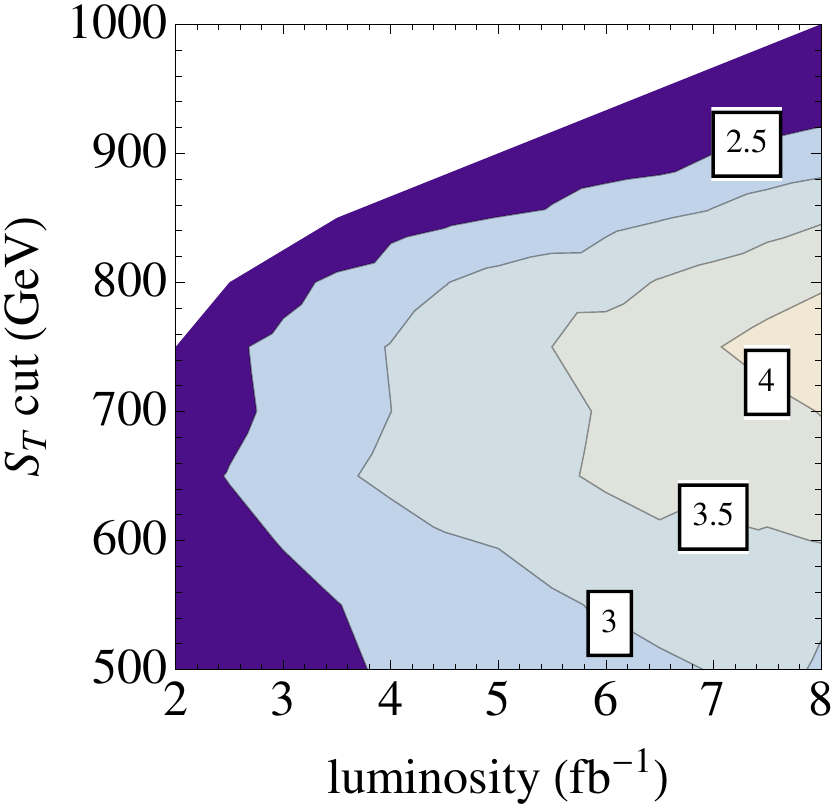}
   \label{wp400g2}
    }\hfill
\caption{Expected significance of the $M_{j1bW}$ variable for a 400
GeV $W'$, as a function of luminosity and $ S_T$ cut.
\label{contourplots1}}
   \end{figure}
When producing these contour plots, we choose a fixed binsize of 50
GeV everywhere except in the upper and lower tails of the
distribution, where we use a superbin. The superbins are sized so that
that no bin ever contains fewer than 50 events.  For higher $S_T$,
there are very few bins between the two superbins, and this makes the
peak-valley-peak structure weak, ruining the significance of the
measurement. Within the white region in the upper left of the plots,
the number of events is so small that no bin with more than 50 events
exists, and our binning strategy gives a null result.  However, for a
high $S_T$ cut one could choose a more sophisticated binning strategy.  
We have verified in a few particular cases that
larger bins for higher $S_T$ cuts can restore some of the significance
of the measurement.  All of this is to say that sophisticated
treatment of the data may lead to a somewhat  better result than our
simple-minded binning strategy would suggest.

   \begin{figure}[h]
    \subfigure[$M_{W'}=600$ GeV, $g_R=2$]{
   \includegraphics[width = \contsize\textwidth] {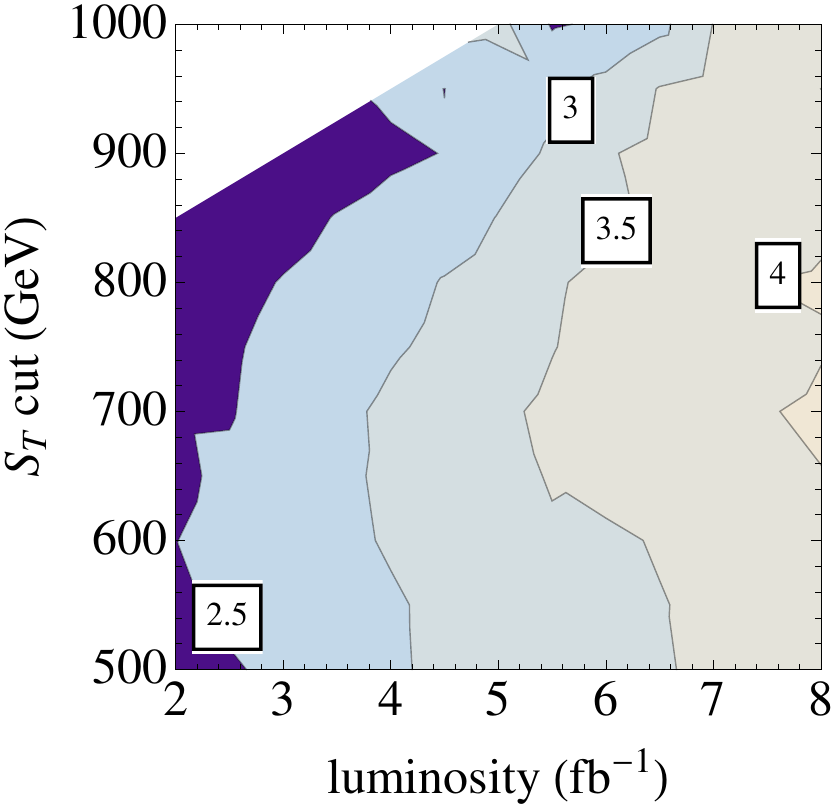}
   \label{wp600g}
    }\hfill
 \subfigure[$M_{W'}=600$ GeV, $g_R=\sqrt{2}$]{
   \includegraphics[width = \contsize\textwidth] {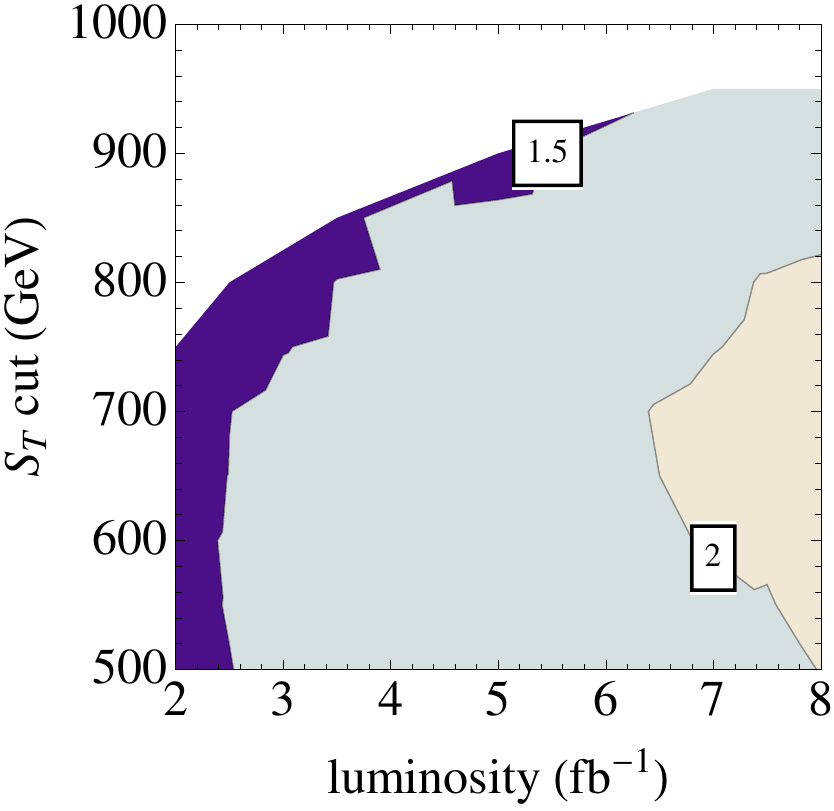}
   \label{wp600g2}}
   \subfigure[$M_{W'}=800$ GeV, $g_R=2$]{
   \includegraphics[width = \contsize\textwidth] {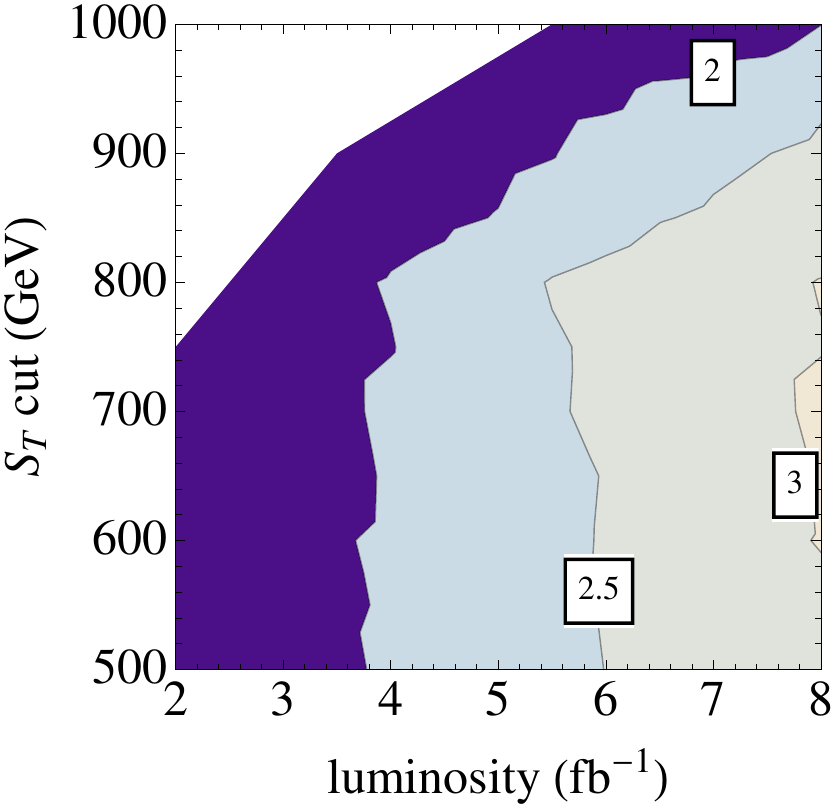}
   \label{wp800g}
    }\hfill
 \subfigure[$M_{W'}=800$ GeV, $g_R=\sqrt{2}$]{
   \includegraphics[width = \contsize\textwidth] {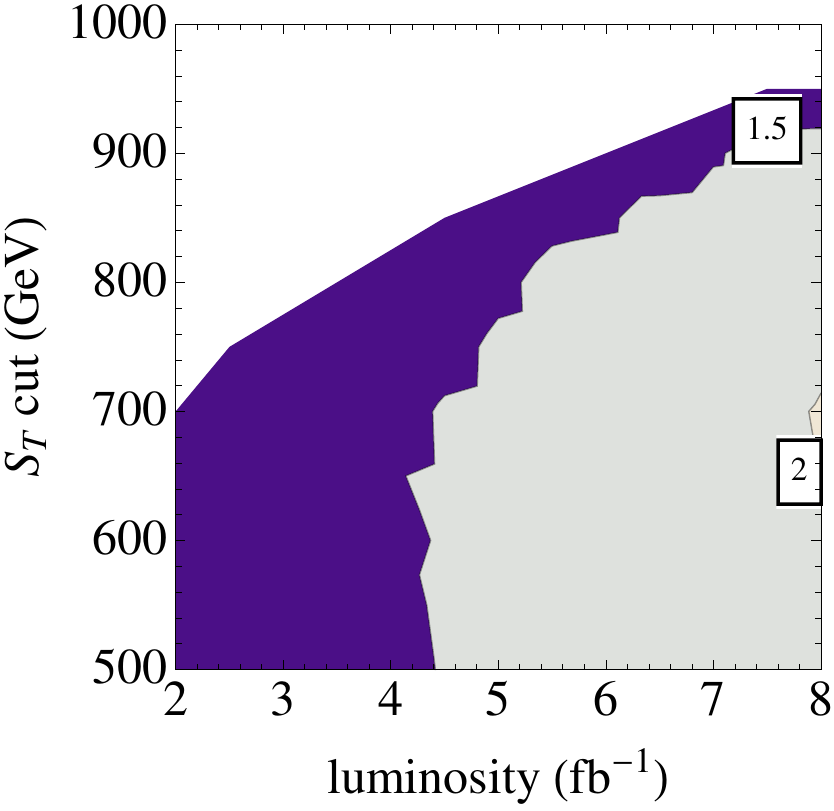}
   \label{wp800g2}
  }
  \caption{Expected significance of the $M_{j1bW}$ variable for a 600 GeV and an 800 GeV $W'$,  as a function of
  luminosity and $ S_T$ cut.\label{contourplots2}}
  \end{figure}

  \clearpage
\subsection{Contour plots for the angle variable}
The plots below show the significance for exclusion of the SM hypothesis using the angle variable, 
along
the lines of our method used for the mass variable.  Note the band of low
significance for the $W'$ with mass of 600 GeV, caused by the shifting
structure that we emphasized in Sec.~\ref{Secangle}; for an $S_T$ cut of around 700
GeV, the asymmetry shifts from one sign to the other.   A study exploiting this dependence of the asymmetry on the $S_T$ cut would have larger significance, but we have not explored this option here.

\begin{figure}[h!]
\subfigure[$M_{W'}=400$ GeV, $g_R=1.5$]{
   \includegraphics[width = \contsize\textwidth] {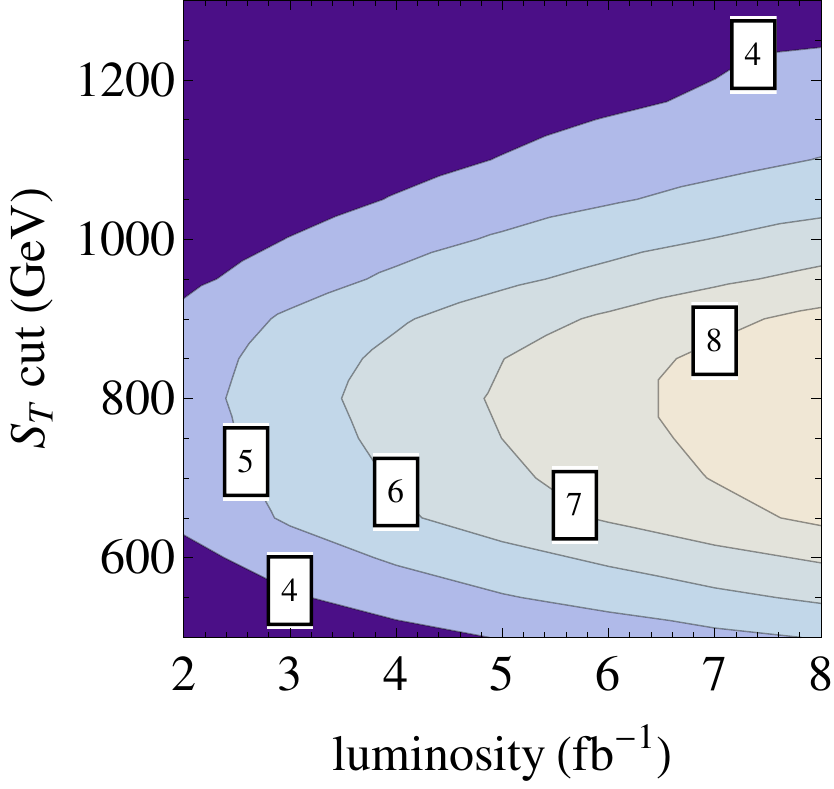}
   \label{wp400g}
 }\hfill
 \subfigure[$M_{W'}=400$ GeV, $g_R=\frac{1.5}{\sqrt{2}}$]{
   \includegraphics[width = \contsize\textwidth] {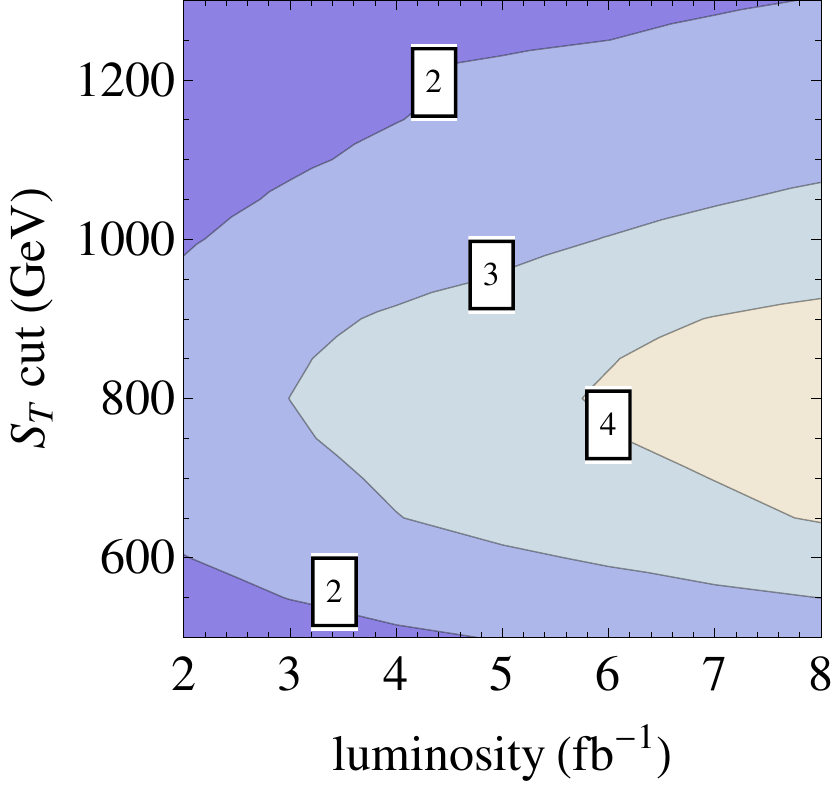}
   \label{wp400g2}
    }
    \subfigure[$M_{W'}=600$ GeV, $g_R=2$]{
   \includegraphics[width = \contsize\textwidth] {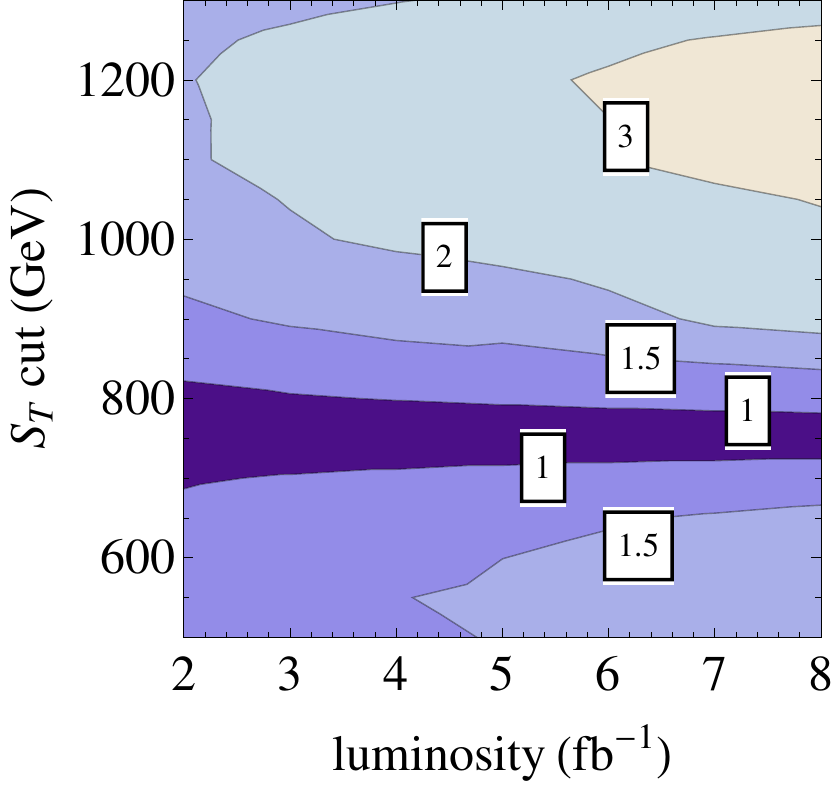}
   \label{wp600g}
    }\hfill
 \subfigure[$M_{W'}=600$ GeV, $g_R=\sqrt{2}$]{
   \includegraphics[width = \contsize\textwidth] {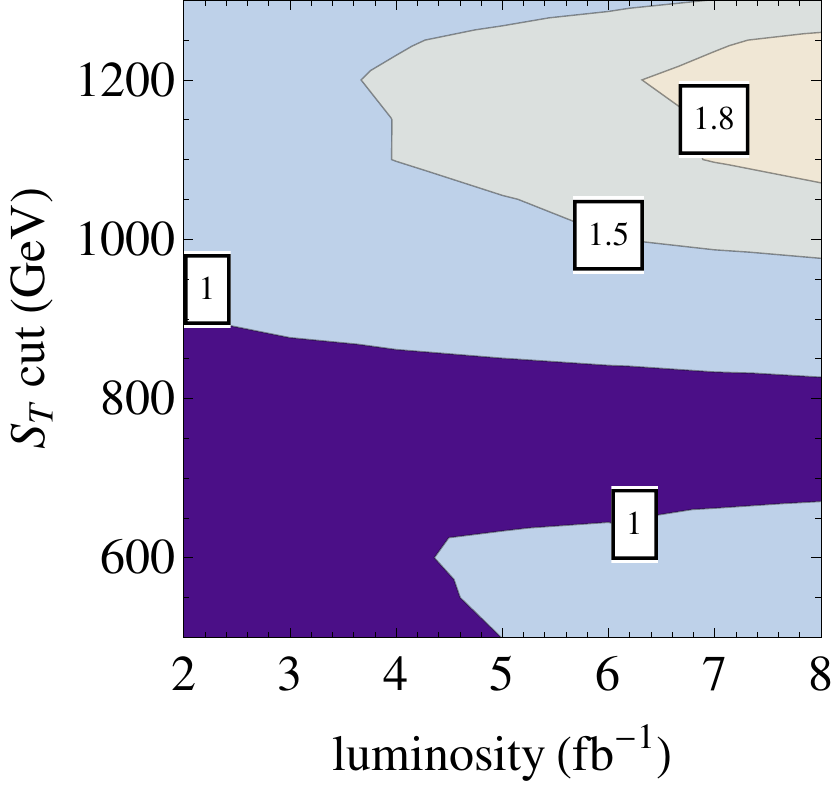}
   \label{wp600g2}}
\caption{Expected significance of $\Delta \phi_{j_1,\ell}$ for a 400 GeV and a
600 GeV $W'$,  as a function of luminosity and $ S_T$ cut. For the 600
GeV $W'$ the dark band corresponds to the range of $S_T$ cuts where
the asymmetry is changing sign, which results in a much reduced
sensitivity. Interference between signal and background is not accounted for.\label{contourplots3}}
   \end{figure}
\clearpage
   \begin{figure}[t]
    \subfigure[$M_{W'}=800$ GeV, $g_R=2$]{
   \includegraphics[width = \contsize\textwidth] {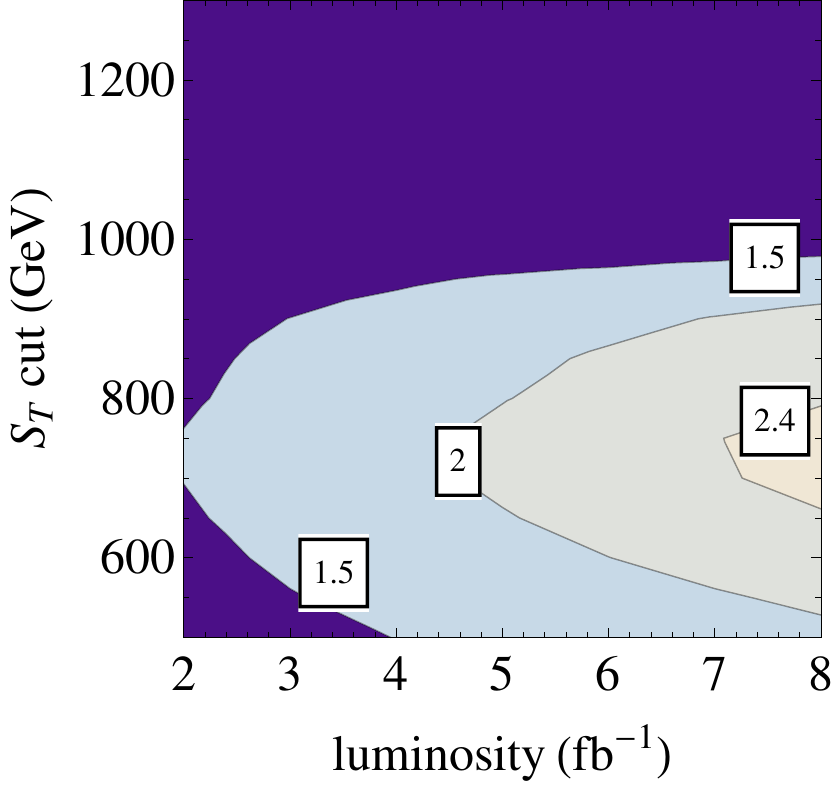}
   \label{wp800g}
    }\hfill
 \subfigure[$M_{W'}=800$ GeV, $g_R=\sqrt{2}$]{
   \includegraphics[width = \contsize\textwidth] {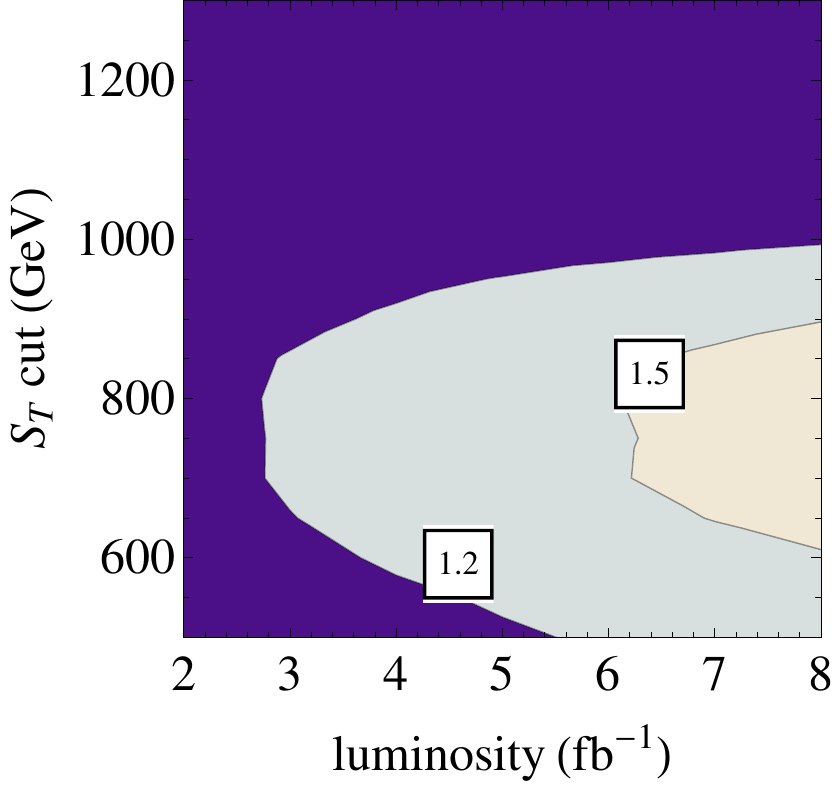}
   \label{wp800g2}
  }
  \caption{Expected significance of $\Delta \phi_{j_1,\ell}$ for an 800 GeV $W'$,
   as a function of luminosity and $ S_T$ cut.   Interference between signal and background is not accounted for.\label{contourplots4}}
  \end{figure}

\section{Strategy Details}
\subsection{Isolation Procedure\label{AppIsolation}}

The detector simulation
DELPHES produces particle candidates and requires the user to
impose the isolation criteria of his or her choice. Hence for each
lepton candidate in the DELPHES output there will be a corresponding
jet candidate, and it is up to the user to decide which one to include in the analysis.
To facilitate this choice, DELPHES provides the user with the following variables for each lepton:
\begin{itemize}
\item$\Sigma P_T$: The sum of the $P_T$ of all the tracks with $P_T > 0.9$ GeV in a
cone of $\Delta R = 0.3$ around the leading
track, excluding that track.
\item $\rho_l$: The sum of the energy deposited in a 3$\times$3 calorimeter grid around
the leading track, divided by
the $P_T$ of that track.
\end{itemize}
Here we lay out the isolation criteria we imposed on the
various particle candidates. An isolated electron is defined as an electron candidate for which $\Sigma P_T<
10$ GeV, $\Sigma P_T< 0.15\:P_{T}^e$ and $\rho_e<1.15$. For isolated muons we require $\Sigma P_T< 10$ GeV, $\Sigma P_T < 0.15\:
P_{T}^\mu$ and $\rho_\mu<0.15$. Finally jet candidates are retained if no isolated leptons are found in a cone of radius 0.3. When a previously isolated lepton is found in a 0.3 cone, the jet candidate is identified with the
lepton and therefore removed from the event. We hereby impose two consistency conditions:

\begin{itemize}
\item No more than 1 isolated lepton is found in a 0.3 cone
\item When one isolated lepton is found, the $P_T$ of the jet candidate can
differ by no more than $10\%$ from
the $P_T$ of the isolated electron.
\end{itemize}
When one of these criteria is not met, we are unable to carry out a consistent
isolation procedure and the entire event is
thrown out. The efficiency of our isolation procedure is $97\%$, both for signal
and background samples.

\section{The $P_T$-difference variables\label{AppPtdiff}}

Among other variables that show charge-asymmetries, ones of possible
further interest include the difference in $P_T$ between the $t$ and the $\bar t$, 
or between the positive and negative $W$ bosons.

Since one top quark is recoiling against the $W'$, while the other top
quark is a decay product of the $W'$, one would expect their
kinematics to differ.  The $P_T$ difference between the $t$ and $\bar
t$ is a variable in which this feature of the signal will manifest
itself.  The same is true for the $W$ bosons from the $t$ and $\bar t$
decays. For each event, we can calculate
 \begin{equation}
\Delta P_{T,W}=\frac{P_{T,W^+}-P_{T,W^-}}{P_{T,W^+}+P_{T,W^-}}\quad\mathrm{and}\quad\Delta P_{T,t}=\frac{P_{T,t}-P_{T,\bar{t}}}{P_{T,t}+P_{T,\bar{t}}}.
\end{equation}
The charge asymmetry at parton-level for these variables can be seen
(for pure signal) in Figs.~\ref{trial1} and \ref{trial3}.

\begin{figure}[h!]
\centering \subfigure[Top $P_T$ difference in signal.]{
   \includegraphics[width=0.45\textwidth] {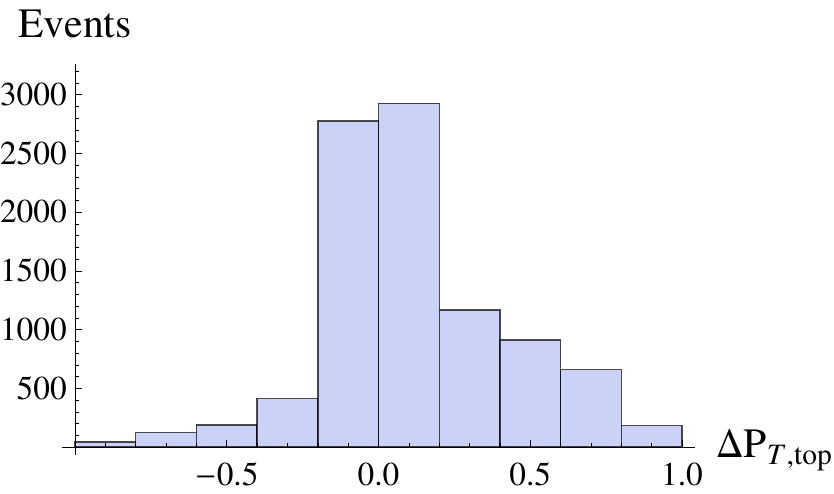}
   \label{trial1}
 }\hfill
 \subfigure[W boson $P_T$ difference in signal.]{
   \includegraphics[width=0.45\textwidth] {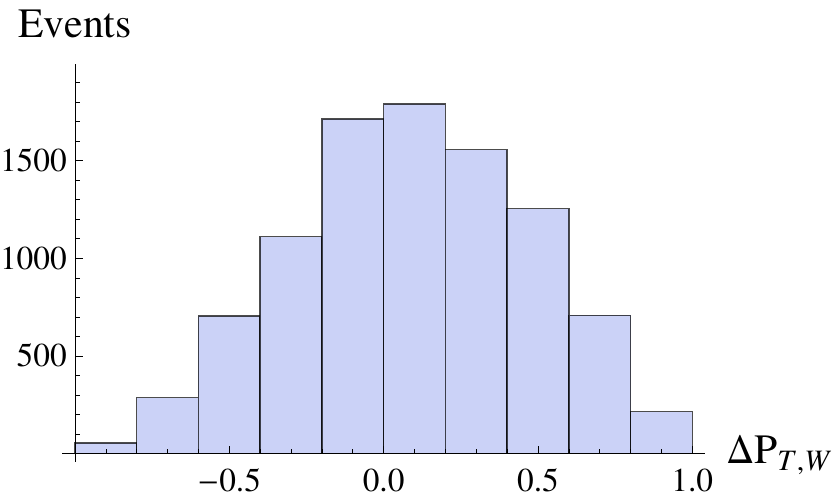}
   \label{trial3}
 }

\caption{Top quark and W boson $P_T$ difference at parton-level in
signal with  a 400 GeV $W'$ with $ S_T$ cut at 700 GeV. The sample is
corresponding to a luminosity of 5 fb$^{-1}$. }
\end{figure}

Although spectacular at parton-level, we find that the $P_T$
difference between the top quarks gets washed out a lot at detector-level by resolution effects and mis-reconstructions. Nevertheless
we encourage experimental colleagues to take this variable in
consideration, since state-of-the-art top reconstruction
methods might alleviate this problem. The $P_T$ difference between
the $W$ bosons is less pronounced at parton-level, but does survive
our detector simulation and the reconstruction of the hadronic $W$.
We find it is particularly useful for a low mass $W'$. Like the angle
variable, it changes sign as a function of the $S_T$ cut.

We have not studied the effect of interference on these
$P_T$ variables. Whether the asymmetry from the SM $t\bar t j$ background is
important also requires further study.

\providecommand{\href}[2]{#2}\begingroup\raggedright\endgroup

\end{document}